\documentclass[journal=jctcce,manuscript=article]{achemso}
\usepackage[version=3]{mhchem} % Formula subscripts using \ce{}
\usepackage{textcomp}
\usepackage{color}
\usepackage{graphicx}% Include figure files
\usepackage{dcolumn}% Align table columns on decimal point
\usepackage{bm}% bold math
\usepackage{multirow}
\usepackage{pdfpages}
\setkeys{acs}{maxauthors = 99}
\usepackage{xcolor}
\usepackage{amsmath}
\usepackage{amssymb}
\usepackage{siunitx}
\usepackage{nameref}
\usepackage{lineno}
\usepackage[colorlinks=true, linkcolor=blue, citecolor=blue, urlcolor=blue]{hyperref}

\newcommand{\fref}[1]{Figure~\ref{#1}}

\author{Ritama Kar}
\affiliation{ 
Department of Chemistry, Indian Institute of Technology Kanpur (IITK), Kanpur - 208016, India%\\This line break forced with \textbackslash\textbackslash
}%

\author{Nisanth N. Nair}
\email{nnair@iitk.ac.in}
\affiliation{ 
Department of Chemistry, Indian Institute of Technology Kanpur  (IITK), Kanpur - 208016, India%\\This line break forced with \textbackslash\textbackslash
}%
\title[An \textsf{achemso} demo]
  {Nature of Hydrated Electron in Varied Solvation Environments}

\abbreviations{IR,NMR,UV}
\begin{document}


\begin{abstract}
%Multiple time step approach could significantly reduce the computational cost of hybrid functional based {\it ab initio} molecular dynamics (AIMD) simulations for condensed matter systems.
%
%In line with this, recently, we proposed two strategies, namely MTACE and s-MTACE, where we combine adaptively compressed exchange (ACE) operator formulation and multiple time step (MTS) integration scheme to reduce the computational cost significantly.
%
%However, these methods suffer from resonance problem, which limits the maximum outer time step that can be used in dynamics. % owing to the transfer of energy between different degrees of freedom.
%
%It restricts 
%
%Utilizing a stochastic resonance free (RF) thermostat with an isokinetic constraint tied to each physical degrees of freedom, we are able to circumvent this issue.
%
Understanding the nature of solvated electrons is important in studying a range of chemical and biological phenomena. 
This study investigates the structural and dynamical behavior of an excess electron in water, examining different solvation environments, including liquid water, ice, monolayer, and chain.
To accurately model these systems, we carry out molecular dynamics (MD) simulations using hybrid density functionals, employing the computationally efficient resonance-free multiple time-stepping based adaptively compressed exchange operator method. 
%instead of the widely used generalized gradient approximation (GGA) based.
%
%However, the high computational cost of hybrid-functional MD has always limited its use for significantly large systems. 
%
%To speed up hybrid functional based MD simulation trajectory here, we employ the developed resonance-free MTACE (RF-MTACE) method.
%
Through these simulations, we create a comprehensive and detailed picture of how excess electrons are solvated across different aqueous environments. 
%
%Our simulations reveal that the
We report the factors influence  the localization and dynamic stability of the hydrated electron.
The determinants include the presence and reorganization flexibility of the dangling OH groups and the spatial arrangement of the surrounding water molecules.
\end{abstract}

%\begin{tocentry}
%\includegraphics[height=4.5cm]{TOC_1st.eps}
%\end{tocentry}


\section{\label{sec:intro}Introduction}
Hart and Boag observed the existence of the hydrated electron by irradiating water with a pulse of 1.8 MeV, producing an absorption band similar to that of an excess electron solvated in liquid ammonia.\cite{ammonia_electron_1864}
Since this observation, the hydrated electron in liquid water has garnered extensive attention, both theoretically and experimentally,\cite{TB_model_JCP_02_HE,turi_Sci_05_HE,Schwartz_LGS_Sci_10_HE,expt_Sci_UV_19_HE,review_PCCP_19_HE,PNAS_theo_expt_23_HE,Schnitker_QCMD_JCP_87_HE,expt_Sci_UV_20_HE,expt_timeResol_photo_JPCL_19_HE,Rg_expt_81_HE,Coe_pes_HE_08}
due to its unique properties and role in chemical and biological processes, such as acting as a simple reducing agent\cite{reactivity_1_HE,reduction_co2_jpcb_20_HE} and DNA damage during radiolysis.\cite{review_radiation_ChemRev_04_HE,DNA_radiation_19_HE}
The hydrated electron also serves as a model system to understand electron localization, solvation, and reactivity in a condensed-matter system.
%

%However, after decades of investigation, 
The direct structural characterization of solvated electron by conventional experiments has been challenging because of its short-lived nature.
Instead, UV spectroscopy\cite{expt_Sci_UV_19_HE,expt_Sci_UV_20_HE}, time-resolved photoelectron spectroscopy\cite{expt_timeResol_photo_JPCL_19_HE}, electron spin resonance (ESR) techniques\cite{Rg_expt_81_HE} have been employed to probe its formation dynamics, localization time scales, and spectral signatures.
These experiments provide geometric information, such as radius of gyration ($r_{\rm g}$), vertical binding energy, through comparison with theory. 
Although, these interpretations critically depend on the accuracy of the theoretical models used to represent electron–water interactions.
Early theoretical and experimental efforts have revealed that the hydrated electron in bulk water exhibits a cavity-shaped structure that is mostly spherical.\cite{TB_model_JCP_02_HE,pp_HE_87,pp_HE_2009,turi_Sci_05_HE} 
However, LGS pseudopotential by Larsen \textit{et al.} challenged this cavity model, sparking considerable debate.\cite{Schwartz_LGS_Sci_10_HE} 
%Despite numerous studies supporting the cavity-like structure of the hydrated electron in liquid water,the development of the LGS pseudopotential by Larsen \textit{et al.} challenged this view, igniting considerable debate.
%Despite of a number of theoretical and experimental observations of cavity like structure of hydrated electron, Schwartz et al. has developed LGS pseudopotential which denied the cavity nature of excess electron in bulk water.
%
Subsequent work using refined pseudopotentials,\cite{comment_1_LGS_11_HE,comment_2_LGS_11_HE,response_LGS_11_HE,schwartz_jpcb_cavity_13_HE} along with path integral molecular dynamics (PIMD)\cite{rahman_PIMD_86_HE,schnitker_PIMD_87_HE,jortner_PIMD_PRL_87_HE,gallicchio_PIMD_dyn_96_HE}and mixed quantum/classical molecular dynamics (QCMD),\cite{TB_model_JCP_02_HE,Schwartz_LGS_Sci_10_HE,Schnitker_QCMD_JCP_87_HE,sprik_QCMD_jcp_88_HE,Schnitker_QCMD_JPC_88_HE,herbert_QCMD_jcp_10_HE,schwartz_QCMD_jctc_16_HE} simulations, helped restore support for the cavity picture, though these methods often relied on parametrized one-electron models with limited predictive power.\cite{review_PCCP_19_HE}

Beyond bulk, the shape and localization of the hydrated electron have been found to depend sensitively on the local hydrogen-bond (H-bond) topology, particularly the presence of dangling OH bonds.
Studies on water clusters have demonstrated the existence of multiple metastable configurations, such as surface-bound or internally localized states, depending on the molecular geometry.\cite{cluster_HE_99,cluster_HE_04,cluster_HE_05,cluster_HE_09}
Turi {\it et al.} demonstrated that even small water clusters could host quasi-cavity-like states stabilized through thermal and structural fluctuations.\cite{turi_Sci_05_HE}
In addition to cluster models, static calculations on crystalline ice, characterized by a rigid H-bond network, have also been used to examine electron solvation, showing localization at lattice cavities or defects.\cite{ice_expt_jpcl_20_HE,ice_pccp_14_HE,ice_pccp_16_HE,ice_surface_prl_05_HE,defect_in_ice}
These studies underscore the significance of local geometry but rely mostly on static models or small systems, raising questions about the dynamical stability of such states under realistic, fluctuating environments.
%Early investigations largely employed mixed quantum/classical approaches, where the electron was treated quantum mechanically while the surrounding water molecules were treated classically.
%
%This methodology,  relied on one-electron pseudopotentials to model electron-water interactions. 
%
%Early work explored the size-dependent stability and electronic structure of anionic water clusters, $({\rm H}_2{\rm O})_n$, revealing how the location and binding strength of the excess electron depend sensitively on the hydrogen bonding topology.
%
%
%Although these approaches yielded insights, their strong parametric dependence often limited their ability to fully explain experimental observations.\cite{review_PCCP_19_HE}
%
%Although PIMD and QCMD approaches addressed critical insights into the temperature dependence and transient structural fluctuations of solvated electron states, they often relied on parametrized one-electron models with limited predictive power.
%
%These shortcomings underscore the need for fully first-principles simulations capable of capturing the electron-water interaction with minimal assumptions, an area where \textit{ab initio} molecular dynamics (AIMD) has become increasingly important.
%

To overcome these limitations, \textit{ab initio} molecular dynamics (AIMD) based on density functional theory (DFT) has become the method of choice to simulate hydrated electrons under thermally evolving, fully flexible conditions.\cite{marx-hutter-book,review_DFT}
Initial AIMD studies of the temperature dependence of excess electron localization\cite{Boero_PRL_03_HE, Boero_JPCA_07_HE} employed generalized gradient approximation (GGA) based functionals, such as BLYP.\cite{PRB_GGA_LYP}
To gain mechanistic insight, cluster-based Born–Oppenheimer molecular dynamics (BOMD) have been extensively used to investigate how H-bond topology influences electron localization, distinguishing between internally and surface-bound states.\cite{PNAS_HGordon_cluster_HE,JPCA_cluster_08_HE,marsalek_cluster_prl_10_HE,marsalek_cluster_jpcb_10_HE,marsalek_cluster_jpcc_10_HE,landan_cluster_jpca_11_HE}
%In 2003, Boero \textit{et al.} first reported Car-Parrinello molecular dynamics (CPMD) simulation with ordinary and supercritical water using BLYP functional\cite{PRB_GGA_LYP}.
%
%Further studies have considered the temperature dependence in solvation of excess electron.\cite{Boero_JPCA_07_HE}
%
%However, 
%Since this method is limited by system size, cluster-based investigations have long provided critical insight into the hydrated electron demonstrating how hydrogen-bond topology governs whether the excess electron localizes internally or on the surface of the cluster.
%
%AIMD simulations on finite water clusters, revealing real-time solvent reorganization and transient cavity formation, thereby highlighting the dynamic nature of electron trapping.\cite{PNAS_HGordon_cluster_HE}
%
Yet, their finite size prevents direct extrapolation to bulk.
As a compromise, hybrid quantum/classical (QM/MM) approaches have been employed, where a few water molecules are treated quantum mechanically (typically with BLYP) and the remaining solvent is described by classical force fields.\cite{QMMM_JPCL_2012_HE,QMMM_JCP_2019_HE,Jungwirth_review_accounts_12_HE,Jungwirth_NatChem_HE_14}
These methods allow for larger systems, but the accuracy is inherently limited by the underlying functional.
%
%Study showed that the nature of the excess electron often becomes indistinguishable from its bulk counterpart, leaving open questions about how dimensionality and interfacial environments influence its structure and stability.\cite{JACS_surface_16_HE}

Notably, 
%across these cluster, QM/MM, and GGA-based AIMD simulations, 
GGA functionals suffer from self-interaction and delocalization errors, leading to incorrect %electron densities
energies,
and electronic structure.\cite{Science_DFT_limitations}
%and underestimation of localization strength.
%
To address the shortcomings, hybrid DFT functionals have been employed, which incorporates a certain percentage of exact exchange.
%to compare the temperature dependence of -based AIMD is better than the commonly used generalized gradient approximation (GGA) based MD.
%\cite{Pasquarello_JPCL_2017_HE,Schwartz_JCTC_22_HE}
%
As reported by Pasquarello and co-workers,\cite{Pasquarello_JPCL_2017_HE} and the recent study by Schwartz and co-workers,\cite{Schwartz_HE_25} dispersion corrected hybrid PBE0 functional with $40\%$ Fock exchange is an optimum choice to describe this system.
These simulations, especially at varying temperatures,\cite{Pasquarello_Angew_22_HE,Schwartz_JCTC_22_HE} show good agreement with experimental observables like absorption spectra and $r_{\rm g}$. 
Furthermore, recent developments in correlated wavefunction methods (e.g., MP2)\cite{Angew_MP2_2019_HE} and machine-learned potentials\cite{venkat_nat_comm_21_HE,Car_PCCP_24_HE} have pushed the frontier of accuracy and sampling, enabling many-body descriptions of the hydrated electron across extended timescales. 
%
%Although, the computational bottleneck to conduct AIMD with hybrid functional is calculating exchange integrl.
%
%Recent advancements in machine-learned models and many-body perturbation theory have further enhanced the accuracy of these simulations, consistently confirming the cavity-like structure of the solvated electron in liquid water.
%Recent advancements with machine-learned models have been utilized in studying hydrated electron with hybrid DFT\cite{Car_PCCP_24_HE} and many-body perturbation theory\cite{Angew_MP2_2019_HE,venkat_nat_comm_21_HE} and correctly reproduce the structural and dynamical properties.
%
%All of the aforementioned studies have confirmed the cavity like structure of solvated electron in bulk water.
%
Nuclear quantum effects (NQE) have been considered by Rybkin \textit{et al.}\cite{venkat_nat_comm_21_HE} which reveals twin-cavity in bulk.

Despite these advances, most hybrid DFT-based AIMD studies have focused on bulk water, leaving unaddressed how different H-bonding topologies affect the structure and dynamics of the excess electron.
Moreover, at the air/water interface, the electron’s structural identity is often ambiguous, appearing indistinguishable from its bulk counterpart.\cite{JACS_surface_16_HE}
This ambiguity leaves open questions about how confinement, reduced dimensionality, and interfacial H-bond arrangements govern localization, and calls for theoretical studies capable of isolating these effects.
By systematically varying the H-bond network environment, 
%from bulk liquid to crystalline ice, defective ice, and reduced-dimensional arrangements, 
we investigate how topological and dynamical features of solvation affect excess electron localization and stability using hybrid density functional.

To this end, we employ the recently developed resonance-free multiple time-step (MTS)\cite{Tuckerman-book} method based on the adaptively compressed exchange (ACE) operator\cite{ACE_Lin,ACE_Lin_1,ACE_2023,JCP_2019_sagar,sagar_JCC_scaling,sagar_JCTC} framework, termed as RF-MTACE.\cite{ritama_jctc}
%We investigate the nature of the excess electron across a range of water solvation environments, with a specific focus on obtaining the response to the structure and dynamics of the excess electron for different H-bonding arrangements of the solvent.
%
%
By effectively decoupling the fast and slow components of forces due to GGA and Hartree-Fock exchange, and by avoiding resonance artifacts arising during the integration equations of motion, 
%that typically limit time-step sizes, 
RF-MTACE achieves a 30–50$\times$ speedup for systems with 
$\sim$100-400 atoms
compared to conventional 
hybrid functional based AIMD simulations.
%velocity-Verlet integration schemes. 
%
This advancement enables long-time and large-scale hybrid DFT simulations that are otherwise computationally demanding and unaffordable. 
%
%Using the RF-MTACE technique, we simulate the excess electron in liquid water, ice (pure and with a molecular defect), and water confined in low-dimensional solvation environments at the level of hybrid DFT.
%

First, we revisit the structural properties of bulk water containing an excess electron.
%and compare our results against established computations and experiments.
%confirming the accuracy of the method in predicting its properties. 
%
We then explore the behavior of the hydrated electron in ice lattice that is perfect and with a molecular defect.
Then we study the nature of the solvated electron when solvating water molecules are arranged in one and two dimensions, as such solvation environments have different orientations and populations of dangling OH bonds.
The dynamic nature of the dangling OH bonds in these systems are also different compared to that of the bulk.
%dynamic behaviour.
%conditions.
%\cite{JACS_surface_16_HE,NatCom_surface_24_HE,NatChem_surface_10_HE}
%
We carried out a detailed examination of electron localization and diffusion behavior. 
We aim to obtain a general picture of electron solvation structure and dynamics, their dependence to different arrangements of dangling OH bonds.
To the best of our knowledge, no previous hybrid AIMD studies have examined the excess electron in such a diverse range of solvation environments.
%making this work a significant step forward in understanding how different H-bonding environments affect electron localization and dynamics. 

\section{\label{sec:methods}Methods}

We performed a series of hybrid functional-based BOMD simulations in the \textit{NVT} ensemble using a modified version of the CPMD code\cite{cpmd,KLOFFEL2021}, which incorporates the RF-MTACE method.\cite{ritama_jctc}
RF-MTACE is a MTS based simulation method.\cite{MTS_1}
Here, the computationally cheap force is computed using an approximate exchange operator within the ACE formalism at every timestep  ($\delta t$) using the initial guess of the wavefunction at that step, while after every $n_{\rm MTS}$ step, i.e., with timestep $\Delta t=n_{\rm MTS}\delta t$, the exact hybrid functional operators are used to compute the slow force.\cite{JCP_2019_sagar}
%will be calculated using the exact exchange operator.
%
In our simulations we chose $\delta t=0.48$~fs and $\Delta t = 48$~ps; i.e., $n_{\rm MTS}=100$.
%has been chosen by modulating the MTS factor $n_{\rm MTS}=100$.
%
%The multiple time step factor ($n_{\rm MTS}$) is chosen as 100 for each simulation
Simulations were performed at 300~K, except for the ice system, which was performed at 77~K.
The temperature was controlled by stochastic iso-kinetic Nos\'{e}-Hoover (RESPA)/SIN(R) thermostat\cite{SINR_main,SINR_20_jctc,SINR_MP_21} with parameters $L=4$, $\tau=9.7$~fs, and $\gamma=0.01$~fs$^{-1}$.
%For optimal efficiency of RF-MTACE, we employed  with a value of 100 for each simulation.
%(details provided in the Supplementary Information).
%
%
We used the PBE0 functional with 40\% exact exchange for the dynamics followed by previous reported studies.\cite{Pasquarello_JPCL_2017_HE,Pasquarello_Angew_22_HE,Schwartz_JCTC_22_HE}
%
%However, we did not include any dispersion correction in our calculation.
%Adaptively compressed exchange (ACE) operator\cite{ACE_Lin,ACE_Lin_1,ACE_2023} was constructed at every 0.48~fs ($\delta t$), while exact exchange calculations were performed at every $n_{\rm MTS}$ step using the MTACE framework.\cite{JCP_2019_sagar,sagar_JCTC,sagar_JCC_scaling}
%
%
%The parameter dependence is thoroughly discussed in Appendix~\ref{AppendixA}.
%
Core electrons were described using norm-conserving Troullier-Martin pseudopotentials,\cite{PRB_TM} generated with GGA functionals, and the plane wave (PW) basis set was expanded with an energy cutoff of 80~Ry.
At each MD step, wavefunctions were converged via direct minimization until the magnitude of the wavefunction gradient became less than $1\times 10^{-6}$ au.
For wavefunction minimization, the direct inversion
of the iterative subspace (DIIS) method\cite{PULAY_DIIS,HUTTER_DIIS} was utilized, and a fifth-order always stable predictor-corrector extrapolation scheme\cite{JCC_ASPC} provided initial guesses for the wavefunctions.
Following the equilibration phase, a negative charge was introduced into each system, and the simulations were run for at least 20~ps (if not mentioned otherwise) with spin multiplicity 2.

We selected 2000 frames, sampled at 10~fs intervals, to analyze the properties of the excess electron.
The center of mass (COM) was calculated for those selected frames as
\begin{equation}
    {\bf r}_{\rm c}=\frac{\sum_{i=1}^{N_{\rm grid}}{\bf r}_i\rho^s({\bf r}_i)}{\sum_{i=1}^{N_{\rm grid}}\rho^s({\bf r}_i)}
\end{equation}
where, $\rho^s=\rho^{\alpha}-\rho^{\beta}$ is the spin density.
The localization of spin density along the MD trajectory was monitored through the $r_{\rm g}$, given by $r_{\rm g}=\sqrt{R_1^2+R_2^2+R_3^2}$, where $R_1^2,R_2^2,R_3^2$ are the eigenvalues of the tensor $\bf{S}$,
\begin{equation}
    \label{eq:rg}
    {\bf S}=\begin{bmatrix}
        \sigma_x^2 & \sigma_x\sigma_y & \sigma_x\sigma_z \\
        \sigma_x\sigma_y & \sigma_y^2 & \sigma_y\sigma_z \\
        \sigma_x\sigma_z & \sigma_y\sigma_z & \sigma_z^2 
    \end{bmatrix} \enspace .
\end{equation}
Here $\sigma_{\bf r}={\sum_{i=1}^{N_{\rm grid}}[{\bf r}_i\rho^{s}({\bf r}_i)-{\bf r}_{\rm c}]}$, for ${\bf r}\equiv x,y,z$.

%
%The visual representations of the systems are shown in the Appendix~\ref{AppendixB}.
%

%\subsection{Bulk water}

To model bulk water, we employed a periodic cubic box containing 64 water molecules, with an edge length of 12.414~\AA.
This setup was also used in previous studies of hydrated electron.\cite{Pasquarello_JPCL_2017_HE,Schwartz_JCTC_22_HE,Pasquarello_Angew_22_HE}

%\subsection{Hexagonal Ice}

%\begin{figure}[h]
%    \centering
 %   \includegraphics[scale=0.5]{Pictures/ice_defect.pdf}
 %   \caption{Hexagonal ice: (a) the perfect lattice (b) with a molecular defect at the position of black arrow; oxygen atoms are shown in red, and hydrogens are in white}
 %   \label{fig:ice}
%\end{figure}

For the ice simulations, we used a hexagonal crystal structure adapted from the work of Pasquarello and co-workers\cite{pasquarello_ice_pre_21}, which follows the Bernal-Fowler ice rule\cite{bernal_fowler_ice_rule} and exhibits P63mc symmetry.
%
%This structure is often called as hexagonal ice with .
%
The system was modeled using a $2\times 2\times 1$ supercell with a periodic orthorhombic box with dimensions $15.64\times 13.55\times 7.36$~\AA $^3$ containing 48 water molecules: See Fig. S1(a), Supporting Information.
The ice was equilibrated for 10~ps using the PBE functional, followed by an additional 2~ps using the PBE0 functional to refine the electronic structure.
In this case, the production run was performed for 10~ps with RF-MTACE. 
%The lattice structures are presented in Figure~\ref{fig:ice}.
%A molecular defect is induced in the perfect lattice, which is shown in Figure~\ref{fig:ice}b.
%The study shows the effect of environmental constraints in stabilising the solvated electron.

%\begin{figure}[ht]
%    \centering
%    \includegraphics[scale=0.6]{layered.pdf}
%    \caption{(a) and (b) Single layer of water molecules with their top and side views, respectively. (c) and (d) Single chain of water molecules with their top and side views, respectively. Oxygen atoms are shown in red, and hydrogens are in white. The dummy carbon atoms introduced to create confinements are represented in black.}
%    \label{fig:layered}
%\end{figure}

%\subsection{Single Layered Water}

To create a monolayer of water, we confined water molecules within two layers of dummy atoms that could add as a repulsive boundary for the water molecules.
For simplicity in setting up this system, we replaced C atoms in two graphene sheets by dummy atoms that has only van der Waals interactions with the water molecules; See Fig.~S1(c), Supporting Information.
The two sheets  were packed along the $X$ direction with a inter-layer gap of 8~\AA. 
The cell dimensions along the $Y$ and $Z$ directions were fixed based on the size of a relaxed graphene sheet, each containing 60 atoms, as in Ref.\cite{Galli_graphene_cnt_wat_08}
%
%Carbon−carbon distances were fully optimized. 
 %
Following the previous studies,\cite{Galli_graphene_cnt_wat_08} $\sim$2~\AA~exclusion zone was maintained at the graphene-water interface.
To achieve a water density of $\sim$1 g/cm$^3$, we used thicknesses of 4~\AA~to accommodate 21 water molecules, the chosen interlayer separation of 8~\AA~was sufficient.
The dummy atoms were treated by molecular mechanics (MM), whereas the water molecules were treated quantum mechanically (QM).
During the simulations, the dummy atoms were kept fixed, and only the interactions between dummy atoms and water were treated by a Lennard-Jones potential.
For this, we employed parameters optimized for the graphene-water system by Deshmukh \textit{et al.}\cite{carbon_wat_vdw_para}
To avoid any artifacts due to periodic boundaries along the surface, we chose a larger cell size of 16~{\AA} 
along the $X$-direction.
After 10~ps of equilibration with the PBE functional, the water molecules formed one layer near the graphene sheets.
%

%This thickness was employed in the case of NT confinement to determine the number of molecules needed to fill up the tubes in order to obtain a density of ∼1 g/cm3.

%\subsection{Water Chain}

To simulate a chain of water molecules, we have placed six water molecules inside a hydrophobic confinement created by a (10, 0) carbon nanotube (CNT) with a diameter of 7.85~\AA~and a length of 12.762~\AA, consisting of 120 carbon atoms.
The carbon atoms of the CNT were replaced by dummy atoms which were interacting with water molecules only through van der Waals interaction, in the same way we modeled a monolayer of water molecules.
%
%We used ($10\times 0$)NT with 7.85~\AA~ diameter containing 120 carbon atoms.
%
As before, 
%with the graphene-water systems,
about 2~{\AA} exclusion zone was maintained between the water molecules and the CNT in the initial structure. 
%
%The nanotube had a length of 12.762~\AA, and six water molecules were placed inside.
%
To mitigate periodic boundary effects, the edge lengths along the $X$ and $Y$ directions were set to 14~{\AA}.
The CNT dummy atoms are treated by MM, whereas the water molecules were treated by QM.
The interactions between dummy atoms and water molecules were also treated as discussed in the previous section.
The system is shown in Fig. S1(d), Supporting Information.

\section{Results}
\subsection{Excess Electron in Liquid Water}

\begin{figure}[ht]
	\centering
		\includegraphics[width=\linewidth]{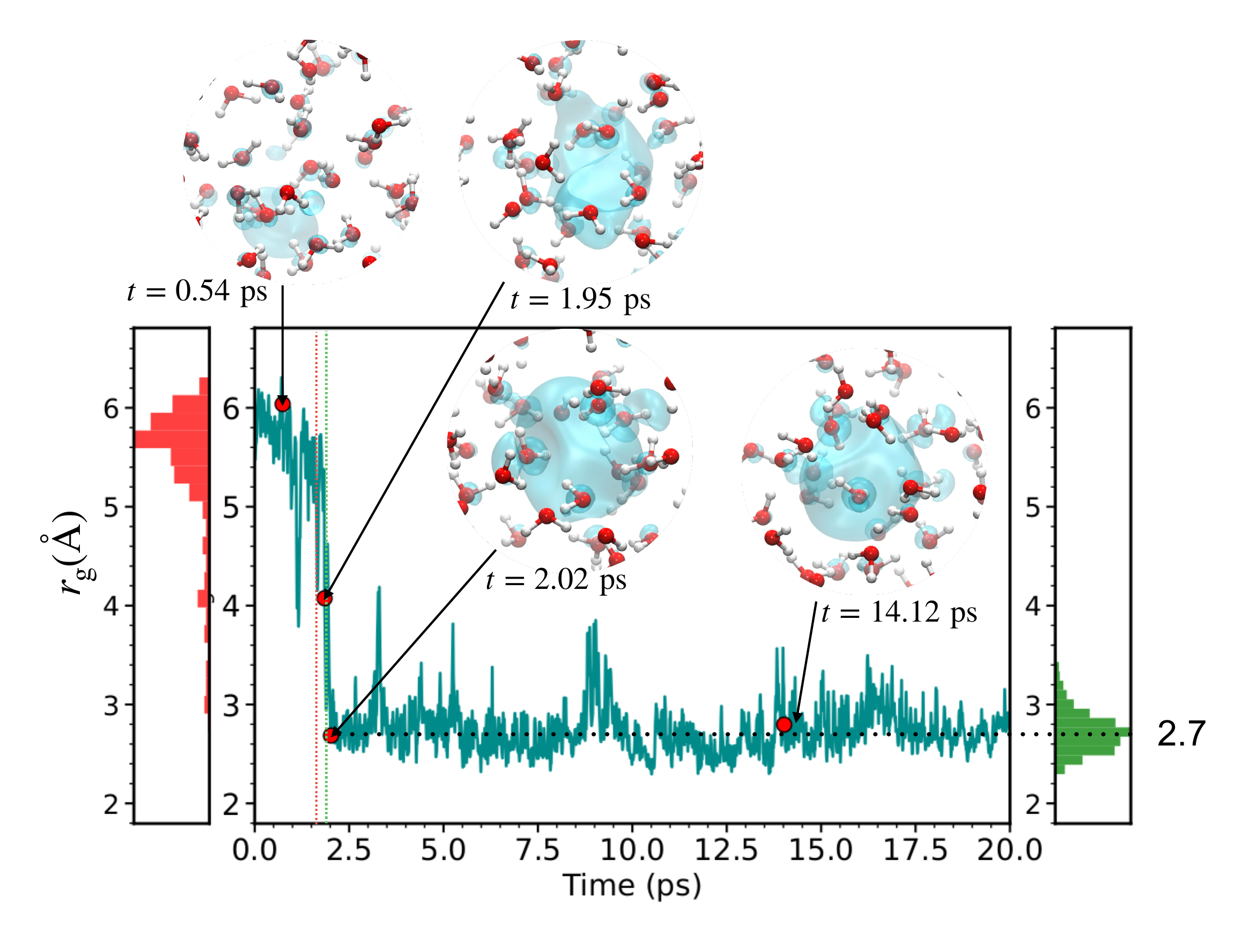}
	    \caption{Time series plot of the $r_{\rm g}$ of the excess $e^-$ in liquid water. Red dots indicate the specific frames for which the corresponding spin density is shown. The histogram is shown separately for the first 2~ps (left panel; red) and after localization (right panel; green). The black dotted line indicates the average $r_{\rm g}$ computed after 2~ps. Spin density plots for selected time frames are shown with an isovalue of 0.001~au. 
        }
	\label{fig:rg_bulk}
\end{figure}

Simulation of excess electron in water was started with an equilibrated liquid water structure without excess electron.
At the onset of the bulk liquid water simulation, the spin density of the excess electron was delocalized but progressively localized within 2~ps.
%into a spherical shape.
This is evident from \fref{fig:rg_bulk}, where we plot the time series data of $r_{\rm g}$.
%during the simulation.
%
%
The histogram of $r_{\rm g}$ values computed before and after localization is also presented.
Initially, the electron exhibits high $r_{\rm g}$ values corresponding to high delocalization, which then stabilized to $r_{\rm g}$ values about 2.7~\AA ~after equilibration.
The equilibrated value lies close to the report from the experimental moment‐analysis of the absorption spectrum yields ($2.48\pm0.1$~\AA) at room temperature,\cite{Rg_expt_81_HE} and transient terahertz spectroscopy ($2.7$~\AA).\cite{Coe_pes_HE_08}
%
%Since these experimental values are not measured directly but inferred via spectral moment analysis, the slight elevation of our average $r_{\rm g}$ is well within the overall error envelope.
%
The pre-solvated structure around 1.95~ps matches with the characteristics of the wet electron having $r_{\rm g}=4.5$~\AA.\cite{Pasquarello_wet_elec}
The average $r_{\rm g}$ value for the equilibrated system matches reasonably well with the {\it ab initio} simulations by Jungwirth {\it et al.}\cite{QMMM_JPCL_2012_HE}
However, our simulation predicts a slightly higher value of $r_{\rm g}$ compared to that by some of the experimental\cite{Rg_expt_81_HE} and the hybrid DFT based simulation data.\cite{Pasquarello_JPCL_2017_HE,Schwartz_JCTC_22_HE,Pasquarello_Angew_22_HE,Schwartz_HE_25} 
%and 
%
The deviation from the previous simulations could be attributed to the choice of the density functional and the basis set; Most of the previous studies used a hybrid PW-Gaussian basis set and  dispersion corrected functionals.

\begin{figure}[ht!]
	\centering
		\includegraphics[width=0.8\textwidth]{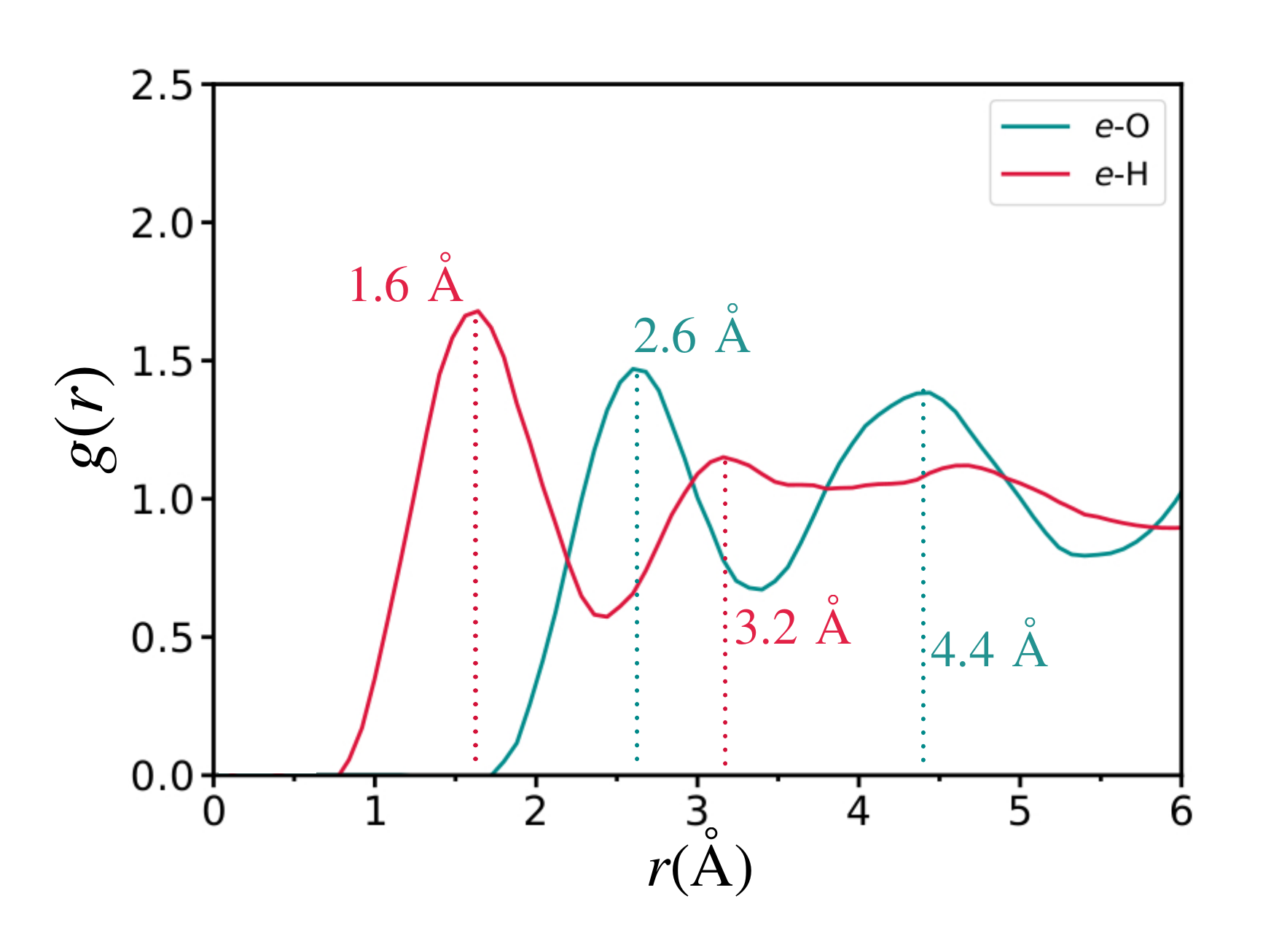}
	    \caption{RDF of oxygen (green) and hydrogen (red) with respect to COM of solvated electron. The peak positions are labeled. (b) Hydrogen atom coordination around COM of solvated electron.
        }
	\label{fig:rdf_bulk}
\end{figure}

To analyze the properties of excess electron in liquid water, we considered only the structures from MD trajectory after the localization of spin density and cavity formation.
The solvation structure of the hydrated electron in liquid water is analyzed through radial distribution function (RDF) calculations. 
%
%The structure of the solvated electron is primarily characterized by the radial distribution function (RDF) between the electron’s COM and the oxygen atoms in the surrounding water molecules. 
%
The electron-oxygen (e-O) RDF, shown in \fref{fig:rdf_bulk}, exhibits a pronounced peak at 2.6~{\AA}, with the first solvation shell extending up to 3.4~{\AA}.
%
%The highest probability of finding oxygen closer to the electronic COM is almost similar to the radius of gyration of the spin density, which infers that the electron is spread upto first solvation shell water molecules.
%
The absence of oxygen atoms within 1.8~{\AA} from the COM of the spin density indicates the formation of a cavity around the electron.\cite{Boero_PRL_03_HE,TB_model_JCP_02_HE,Schnitker_QCMD_JCP_87_HE,Schnitker_QCMD_JPC_88_HE} 
%
%The RDF for e-O shows a prominent peak at 2.4 \AA~, with the first minimum at 3.2 \AA~, defining the electron’s first solvation shell. 
%
This structured solvation shell suggests stabilization of the electron by surrounding water molecules. 
The second solvation shell shows a similar arrangement, highlighting its significance in cavity stabilization.
Since the average $r_{\rm g}$ is close to the first maximum of the e-O RDF, we can infer that the excess electron is primarily confined within the first solvation shell.
The electron–hydrogen (e–H) RDF starts at 0.8~{\AA} distance from the COM of spin density, with the highest peak at 1.6~{\AA}, confirming the orientation of hydrogen atoms towards the cavity.
%hydrogen atoms orient towards the e-COM, stabilizing the electron via a single O-H moiety. 
%

%

\begin{figure}[ht!]
	\centering
		\includegraphics[width=0.8\textwidth]{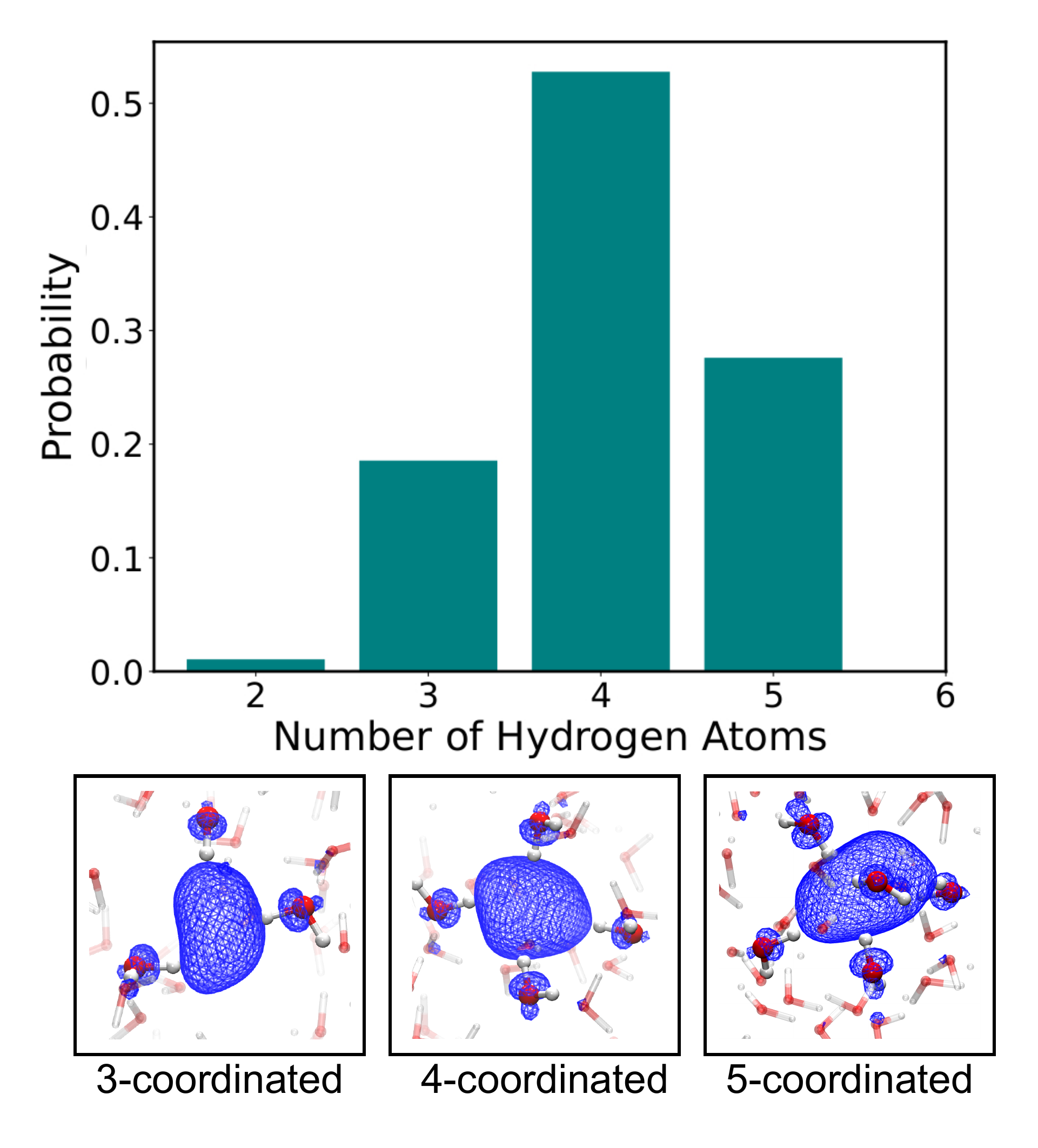}
	    \caption{Hydrogen atom coordination around COM of solvated electron. Spin densities for three different structures of coordination are also displayed. The interacting water molecules are shown with solid spheres and bonds, whereas others are shown in transparent. Spin densities are plotted with mesh for isovalue 0.001~a.u.
        }
	\label{fig:coord}
\end{figure}

Coordination number analysis was performed considering water molecules were selected based on oxygen atoms located within 3.4~{\AA} and hydrogen atoms within 2.4~{\AA } of the COM of the electron (\fref{fig:coord}).
We observed that the most probable configuration involved four hydrogen atoms within the first solvation shell.\cite{Schwartz_HE_25}
Configurations with three or five hydrogen atoms also occur with significant probability, reflecting the fluctuation of H-bond network.\cite{Boero_PRL_03_HE,Boero_JPCA_07_HE,Pasquarello_JPCL_2017_HE,Pasquarello_Angew_22_HE,Car_PCCP_24_HE,Angew_MP2_2019_HE,Schwartz_JCTC_22_HE}
On average, a near-spherical hydration shell
%(asphericity measurement in Fig. S6, Supporting Information) 
formed around the electron (as confirmed from the inset images; also in Section S2 and S4, Supporting Information), corroborating earlier studies.\cite{Schwartz_JCTC_22_HE,Pasquarello_Angew_22_HE}
Interestingly, three- and four-hydrogen coordinated structures exhibited slightly higher $r_{\rm g}$, whereas higher coordination numbers correlated with more compact electron localization.\cite{venkat_nat_comm_21_HE}
This explains our observation of slightly elevated $r_{\rm g}$ compared to computations.\cite{Pasquarello_Angew_22_HE,Car_PCCP_24_HE,Angew_MP2_2019_HE,Schwartz_JCTC_22_HE}

%As shown by Rybkin {\it et al.}, excess $e^-$ coordination by three or four water molecules will have slightly higher $r_{\rm g}$, whereas higher coordination leads to more compact localization and lower $r_{\rm g}$.
%

%

\begin{figure}[ht!]
	\centering
		\includegraphics[width=\textwidth]{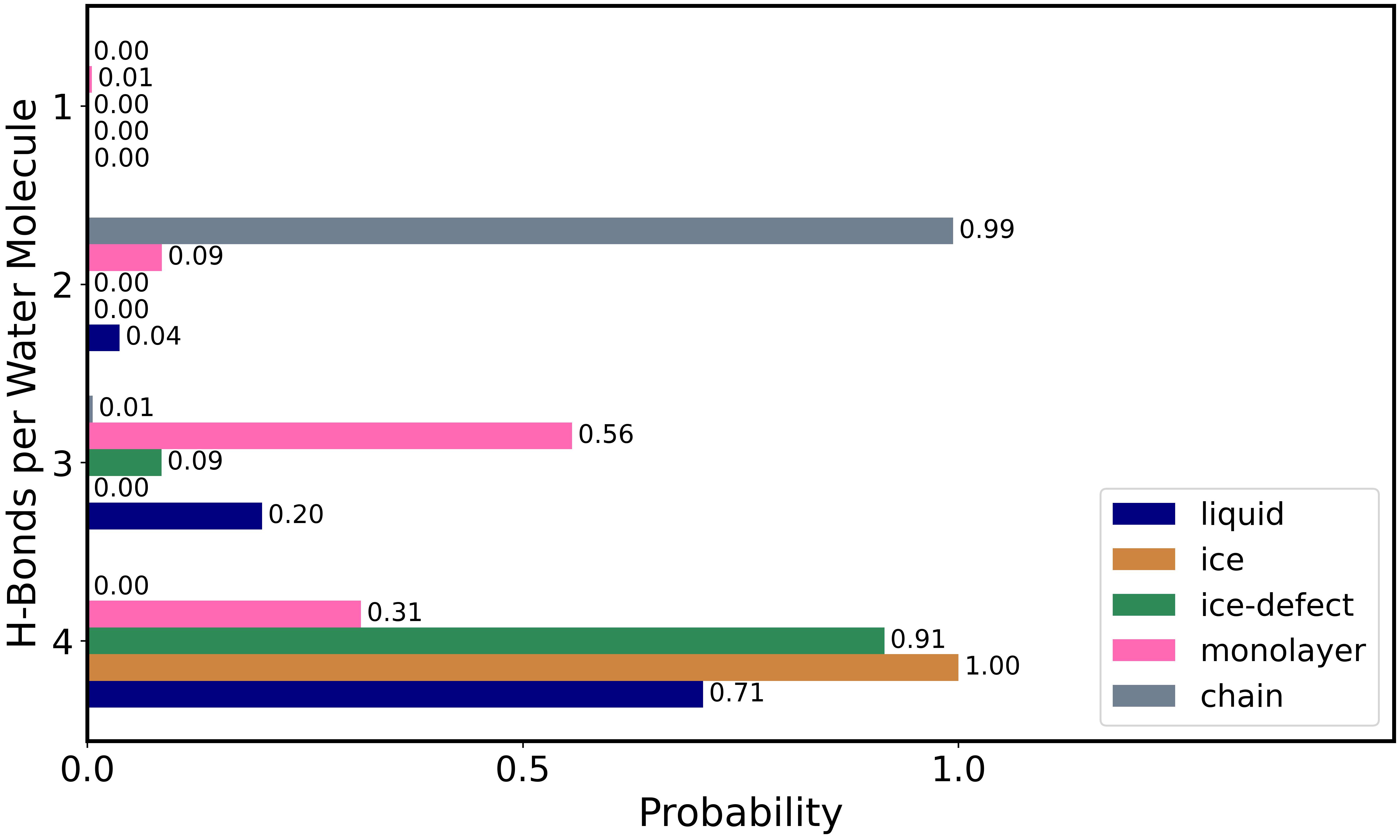}
	    \caption{Average H-bond distribution per water molecule for different systems.
        }
	\label{fig:hbond_all}
\end{figure}

%\begin{figure}[htbp!]
%	\centering
%		\includegraphics[width=0.9\textwidth]{HeH_angle.pdf}
%	    \caption{Anglular distribution of H-$e$-H for various coordination numbers is presented, with the maximum probable values labelled; a graphical illustration of the angle for four coordinated $e^-$ is in the inset. 
%        }
%	\label{fig:heh}
%\end{figure}

%We further investigated the angular distribution of hydrogen atoms around the electron.
%
%Specifically, we analyzed the H–e–H angle distribution for different coordination numbers, as shown in \fref{fig:heh}.
%
%The inset of \fref{fig:heh} illustrates a four-hydrogen configuration, which shows the maximum peak at $\sim$99\textdegree, roughly resembling a near-tetrahedral geometry centered on the electron.
%
%A three-hydrogen coordination peaks at $\sim$92\textdegree,indicative of a trigonal arrangement, while five-coordination results in a peak at  $\sim$116\textdegree~, consistent with a trigonal-bipyramidal geometry. 
%

%
%the roughly spherical organization of water molecules within the first solvation shell, characterized by distinct angular peaks

\begin{figure}[ht!]
	\centering
		\includegraphics[width=0.8\textwidth]{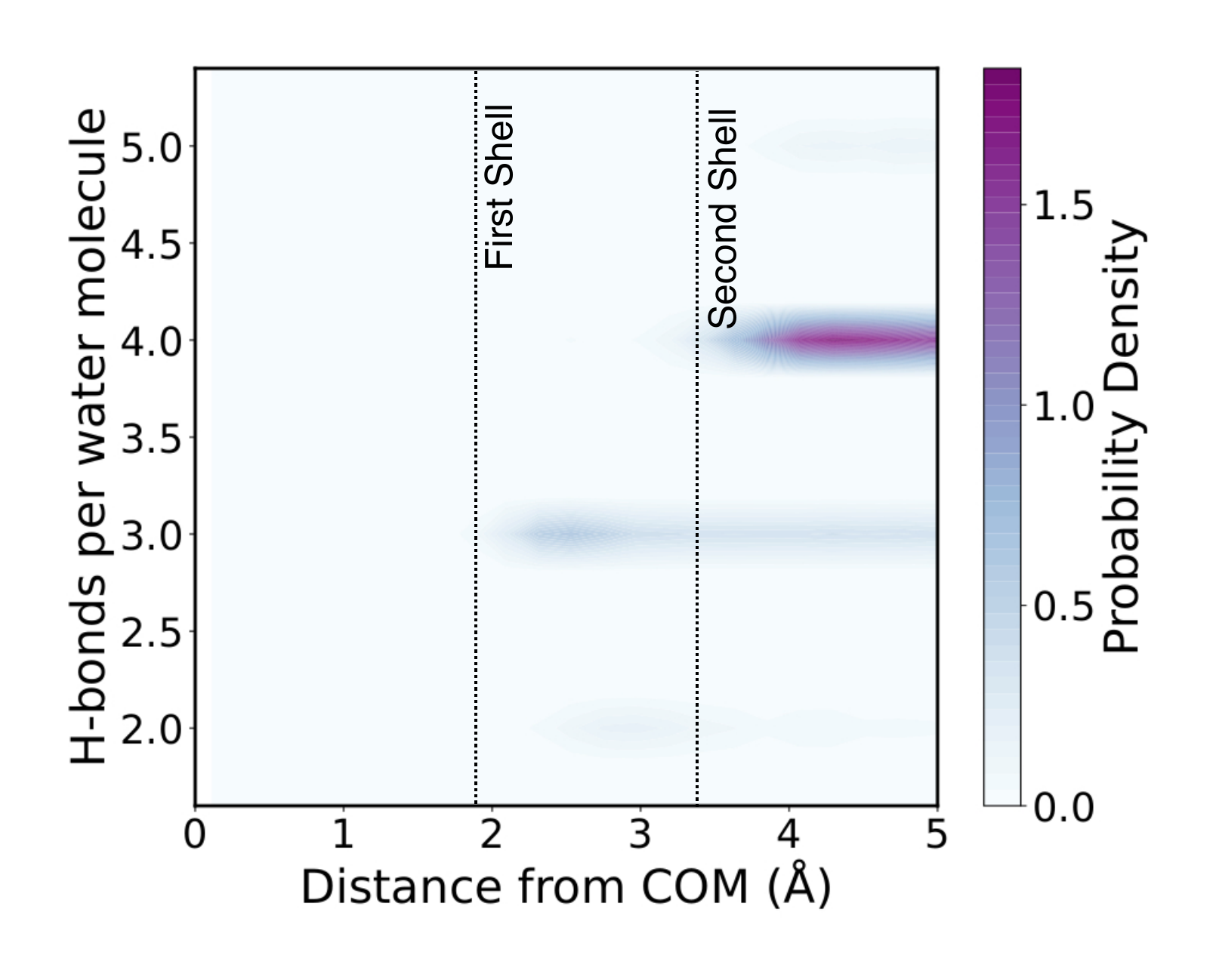}
	    \caption{Average H-bond distribution per water molecule along the distance from the COM of spin density in liquid water.
        }
	\label{fig:hbond_bulk}
\end{figure}

In pure liquid water, each water molecule typically forms four H-bonds, resulting in a tetrahedral coordination.
However, the presence of the excess electron perturbs this structure significantly. 
Only $70\%$ of water molecules maintained tetrahedral hydrogen-bonding, and the probability of three-coordinated water molecules rose to $\sim 20\%$ (Figure~\ref{fig:hbond_all}, detail in Section S3, Supporting Information).
%It has been observed that there is a significant reduction of four H-bonded water molecules present in liquid water, arising  probability of three H-bonded water molecules .
%
To quantify the spatial extent of this perturbation, we calculated the average number of H-bonds per water molecule as a function of distance from the electron COM.
\fref{fig:hbond_bulk} shows that no H-bonds are formed within 1.8~{\AA} from COM, reinforcing the existence of a cavity.
Between $\sim$1.8–3.4~{\AA} (first shell), the average dropped to 3, while beyond 3.4~{\AA} the bulk-like value of 4 was recovered.
%Between $\sim$1.8–3.4~\AA, corresponding to the first solvation shell, the average number of H-bonds per water molecule appears to be three.
%
%Beyond this shell, the coordination number returns to four per water molecule, indicating the re-establishment of bulk-like tetrahedral order.
%
The under-coordination of first solvation shell water molecules was also seen in previous studies.\cite{Pasquarello_Angew_22_HE,Car_PCCP_24_HE,Angew_MP2_2019_HE,Schwartz_JCTC_22_HE,venkat_nat_comm_21_HE}
%This transition marks the emergence of a bulk-like second solvation shell, a structural feature that serves as a cage around the cavity.
%
%While the first shell provides the deformable region necessary for cavity formation through broken or weakened hydrogen bonds, the second shell contributes mechanical rigidity and stabilization. (discussion)
%We also analyzed the types of H-bond interactions within the first solvation shell. 
%
%As shown in \fref{fig:hbond_bulk}b, the most common bonding configuration is two acceptors and one donor H-bond (2A1D), with a marked decrease in the 2A2D configuration typically seen in bulk water.
%
%This reduction in donor H-bonds is attributed to the presence of dangling OH as a result of their interaction with the excess electron.
%

\begin{figure}[htbp!]
	\centering
		\includegraphics[width=0.9\textwidth]{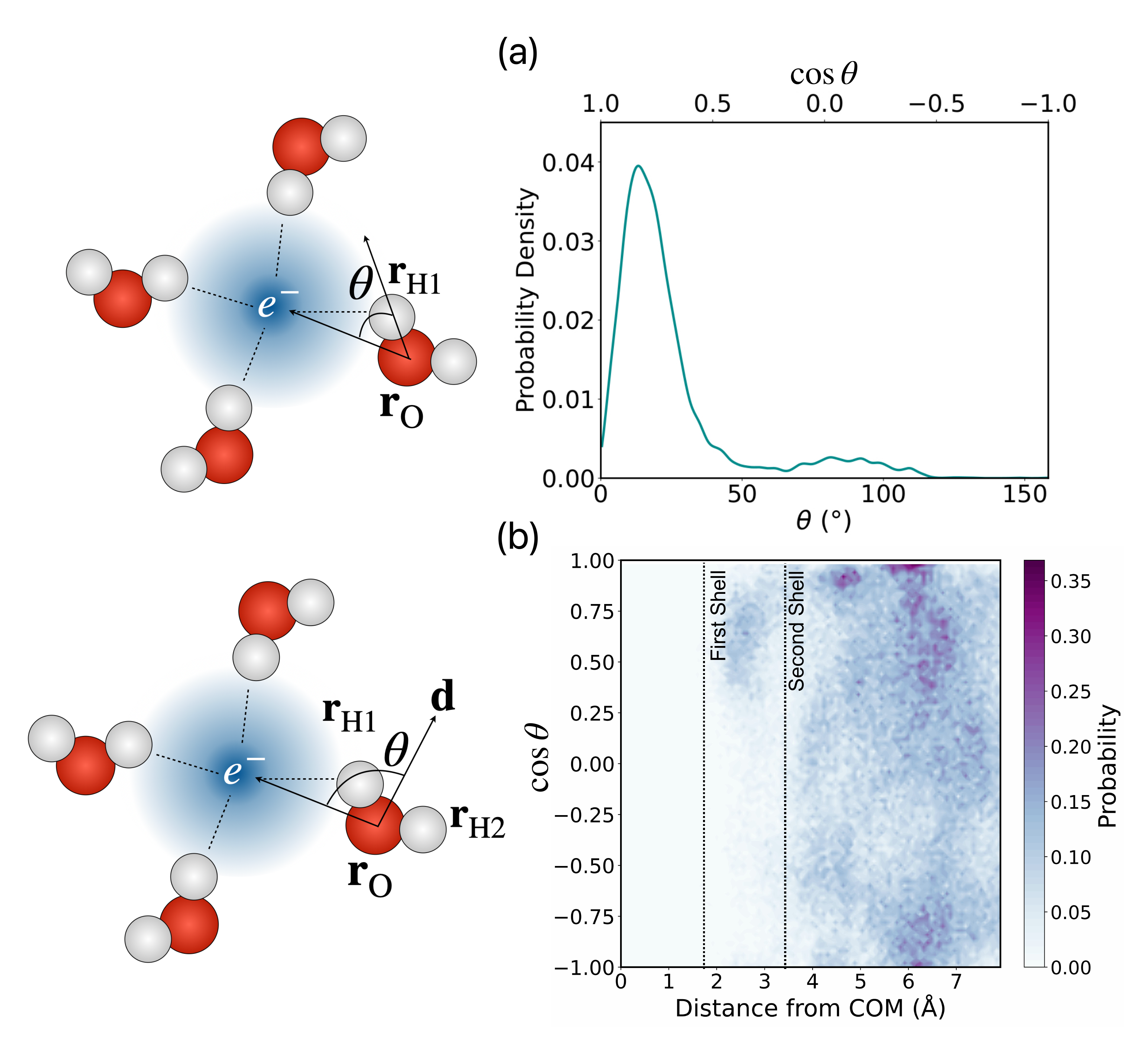}
	    \caption{(a) Average angular distribution of COM of the electron to O-H vectors of the first solvation shell water molecules; (b) Histogram of cosine angle between dipole vector of water and its direction to COM is presented with varying the distance. The dipole is calculated as ${\bf d}=({\bf r}_{\rm H1}+{\bf r}_{\rm H2})/2-{\bf r}_{\rm O}$. The angles are shown in the illustration. 
        }
	\label{fig:eoh_dip}
\end{figure}

Orientational ordering of water molecules around the excess electron was analyzed using the angular distributions of O–H bonds and water dipole vectors relative to the vector connecting the oxygen atom to the COM (\fref{fig:eoh_dip}).
In the first solvation shell, a strong preference was observed for O–H bonds to point toward the electron at angle less than $\sim$30\textdegree, see Figure~\ref{fig:eoh_dip}a, consistent with a hydrogen-bond-like interaction.\cite{venkat_nat_comm_21_HE,Car_PCCP_24_HE}
%Most of the O–H vectors are oriented at angles less than  toward the electron cavity. 
%
%It suggests that electron stays in the cavity with an H-bonding-like interaction with the first solvation shell water molecules, in agreement with prior computational studies.
%
We also examined the average distribution of dipole vector orientations of every water molecule with its distance from the COM (\fref{fig:eoh_dip}b).
Specifically, we plotted the distribution of the cosine of the angle, which is defined between the dipole moment vector of a water molecule and the vector pointing to the COM.
We found the dipole vectors of first-shell water molecules point inward, corresponding to angles of approximately lower than $60$\textdegree, i.e., creating an electropositive pocket, with $\cos\theta$ peaking between 0.5 and 0.75.\cite{Schwartz_HE_25}
Such an orientation ensures that only one hydrogen per molecule interacts with the electron, while the other hydrogen points away and remains engaged with the bulk water network.
%
%This apparent contradiction indicates that only one hydrogen from each water molecule is actually oriented toward the electron and involved in a hydrogen-bond–like interaction, 
%
This joint angular distribution of O–H vectors and dipoles provides  a perspective on how the orientation of water molecules in the first solvation shell stabilize the excess electron.
%This behavior complements our previous observation of hydrogen atoms aligning toward the cavity, suggesting a compromise between maximizing favorable {electrostatic} interactions and maintaining the H-bonding network.
%
Beyond the first shell, both bond and dipole orientations became isotropic, spanning the full range of $\cos\theta$ values (from –1 to 1).
Thus it is clear that the  structural reorganization induced by the excess electron is localized.
%
%Taken together, these findings confirm that the presence of the excess electron not only disrupts the H-bonding topology but also imposes a distinct orientational ordering on nearby water molecules, which is gradually lost beyond the first solvation shell.

\subsection{Excess Electron in Ice}

\begin{figure}[ht]
	\centering
		\includegraphics[width=\linewidth]{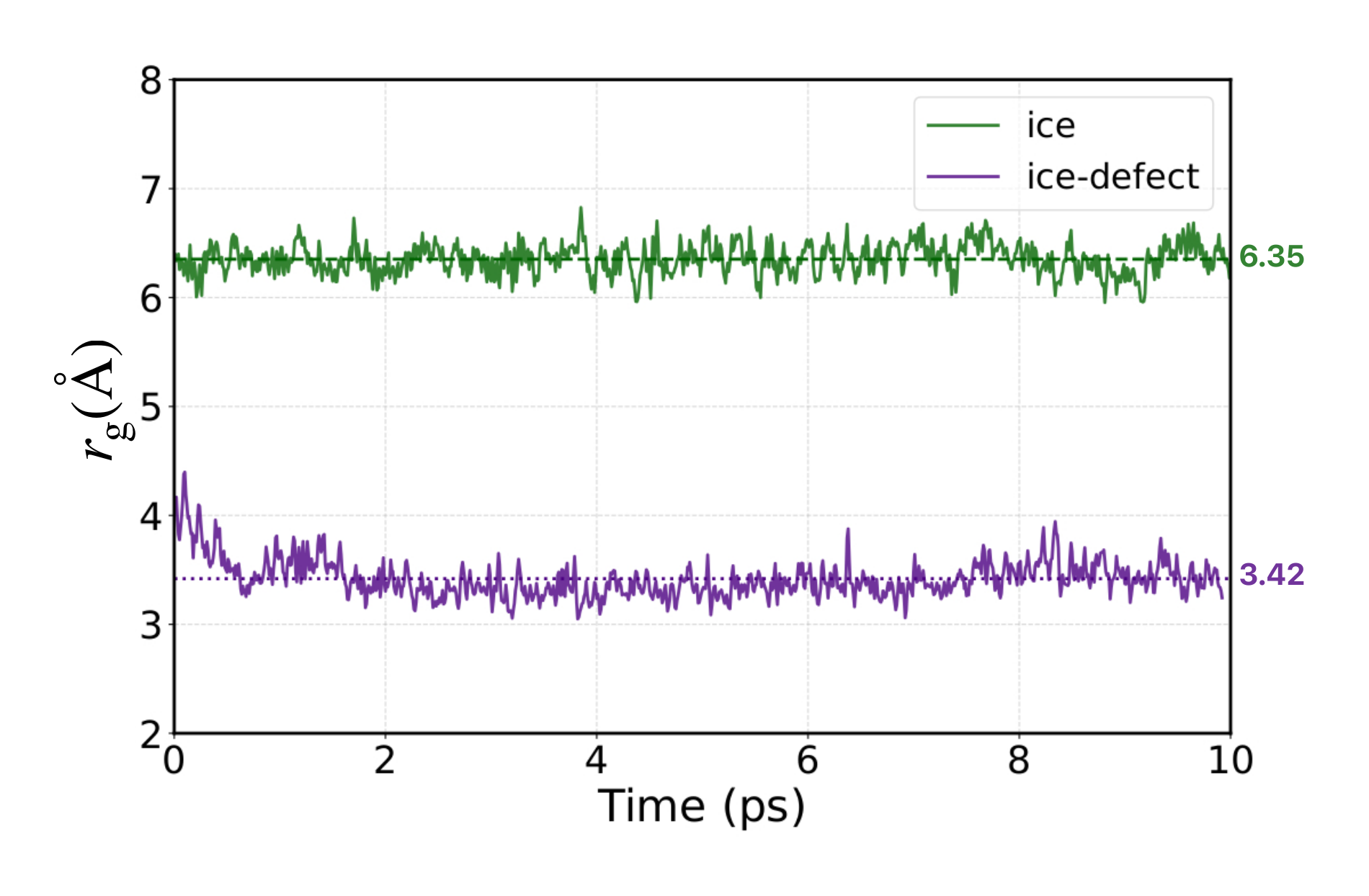}
	    \caption{Time series plot of the $r_{\rm g}$ of the excess $e^-$ in ice, with and without defect.}
	\label{fig:rg_ice}
\end{figure}

\begin{figure}[htbp!]
	\centering
		\includegraphics[width=\textwidth]{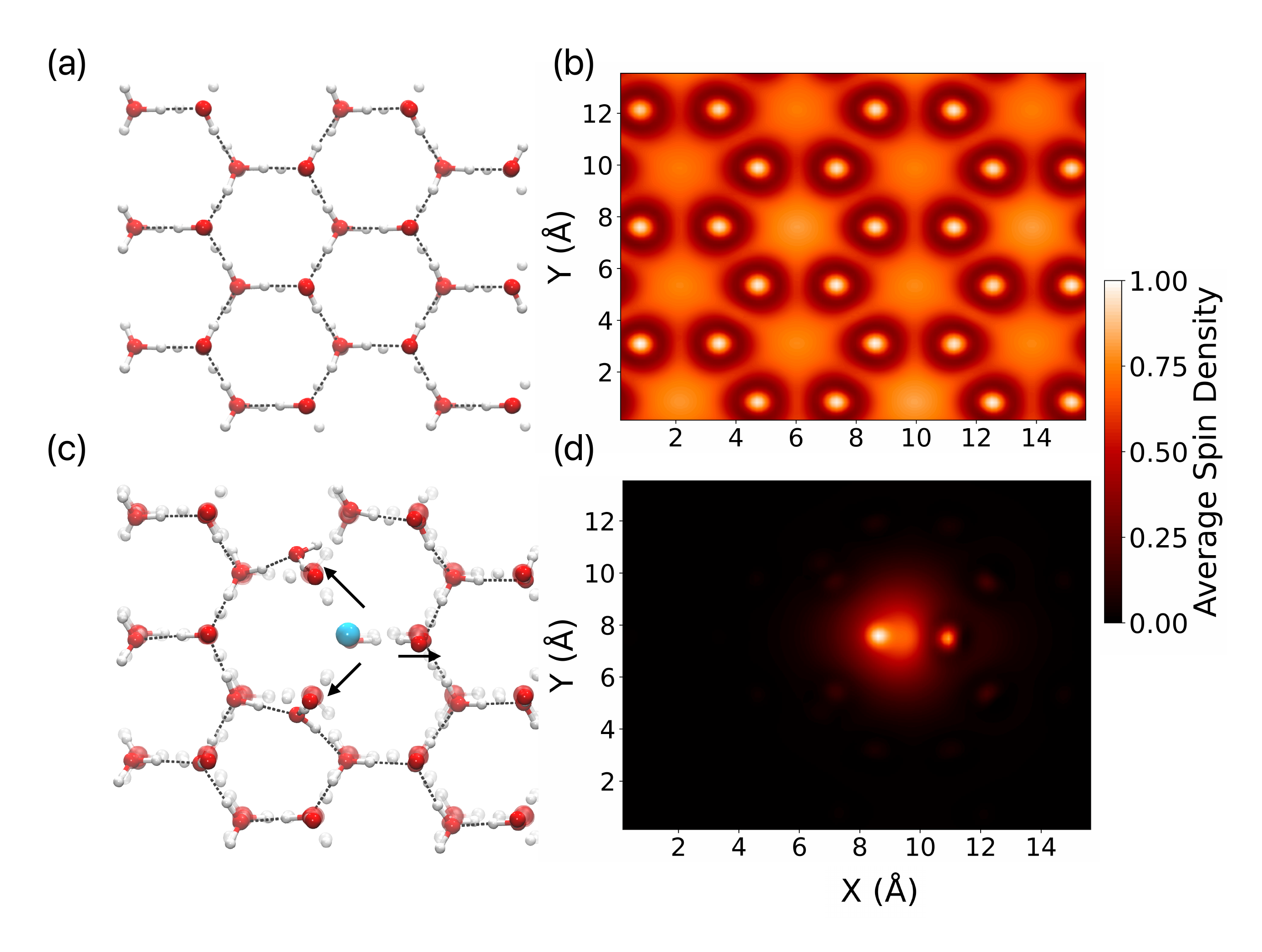}
	    \caption{(a) Snapshot from MD simulation of perfect hexagonal ice with an excess electron, H-bonds are shown by black dashed lines; (b) Time-averaged 2D projection of the normalized spin density for the structure in (a), maintaining the same orientation; Snapshot of ice containing a molecular defect and an excess electron is shown in (c); The vacancy location is marked with a cyan sphere, transparent atoms show the original lattice without the excess $e^-$. Direction of molecular displacements due to the defect are illustrated with black arrows; (d) Averaged 2D spin density projection of the defective ice system maintaining the same orientation as (c).
        }
	\label{fig:proj_hb_ice}
\end{figure}
Water molecules and the hydrogen bond network could reorganize in liquid water to stabilize the excess electron by the formation of a cavity.
This observation encouraged us to explore a solvation environment where the reorganization of hydrogen bonds is not viable or hindered.
In particular, we studied the solvation behavior of the excess electron within hexagonal ice at ~77~K.
For the case of the ideal crystalline structure,  the average $r_{\rm g}$ value (6.35~{\AA}) was found to be high (\fref{fig:rg_ice}), implying a delocalized character.
The spin-density remained delocalized in the lattice throughout the dynamics.
%
%Time evolution analysis of the (Figure~\ref{fig:rg_proj_ice}a) in perfect ice reveals that $r_{\rm g}$ remains stable at a high value (~6.35~\AA) throughout the simulation, indicating that the electron remains completely delocalized across the rigid ice lattice with no significant deviation over time. 
%
Such a behavior was expected considering the rigidity of the lattice, which restricts reorganization of hydrogen bond network, thereby inhibiting the cavity formation.
%This consistent high value further emphasizes the inability of the ideal crystalline ice structure to localize the excess electron.
%
Projected spin density analysis in \fref{fig:proj_hb_ice}a-b showed bright regions localized on lattice voids and oxygen atoms, but no stable localization, indicating that the electron transiently occupies pre-existing interstitial sites.
This is clear, as all of the water molecules are four-coordinated and there is an absence of  dangling OH groups that are necessary to form stable solvation; See \fref{fig:hbond_all}.
%as confirmed from average H-bond analysis for the system in 
%which 
%This delocalized behavior is consistent with the absence of fast structural fluctuations in ice at low temperatures.
Small amplitude oscillations of $r_{\rm g}$ for ice implicate that the delocalized nature of the spin-density remains as such throughout the dynamics.
%This delocalized state is further corroborated by the projected spin density, which shows bright regions centered on oxygen atoms and pre-existing lattice cavities.
%
%These regions of increased electron density highlight the absence of any stable localization site within the ideal ice lattice, reinforcing the delocalized nature of the excess electron.
%We observed negligible structural relaxation when comparing ice containing an excess electron to pure ice without the electron (see ).
%

However, crystalline ice often contains various defects.\cite{defect_in_ice}
Among these, we have adapted a molecular point defect, which is the simplest and relevant for our study; See Fig. S1(b), Supporting Information.
Such a defect, formed by the removal of a lattice water molecule,  disrupts the hydrogen-bonding network by removing two donor and two acceptor hydrogen bonds, leaving behind dangling O–H groups, as shown in \fref{fig:hbond_all}.
%
%To probe how such a defect influences electron localization, we introduced a vacancy and monitored the behavior of the system across the MD trajectory.
%
\fref{fig:rg_ice} revealed that the $r_{\rm g}$ fluctuated around $3.4$~{\AA} throughout the dynamics, suggesting that the electron has a localized structure. 
Notably, no deviation in $r_{\rm g}$, or large amplitude fluctuation was observed over time, which signifies that the electron localization at the defect is stable. 
The snapshot from the MD simulation in \fref{fig:proj_hb_ice}c reveals structural relaxation around the defect.
However, this relaxation is purely translational in nature. 
The local environment adjusts, although,  reorganization of the water molecules was absent.
%and disturbance to the surrounding bulk structure. 
%
The constraint is due to the rigidity of the ice lattice at low temperatures inhibiting any substantial reorientation.
The projected spin density (\fref{fig:proj_hb_ice}d) shows that the excess electron is now concentrated at the vacancy site, as opposed to its delocalized nature in perfect ice. 
Our observation is consistent with the previous report by Antonelli {\it et al.}\cite{ice_pccp_16_HE} who performed structural optimizations of the system with PBE0.
These results underscore the crucial role 
%of structural defects, specifically 
%of reorganization of solvating water molecules to provide sufficient dangling OH groups
of under-coordinated water molecules with dangling OH groups 
in facilitating electron localization.
%in the otherwise rigid and delocalized ice lattice.
%
%

%As presented in \fref{fig:hbond_all}, perfect ice maintains full tetrahedral coordination, with each molecule forming four hydrogen bonds.
%
%The stable H-bonding network supports a more delocalized electron distribution in ice.
%
%Upon introducing the molecular defect, however, the tetrahedral geometry is locally broken, reducing the percentage of four H-bonded water molecules present in the lattice.
%
%This creates a favorable environment for electron localization.
%

% 
%As the dynamics go, 
%This under-coordination, particularly in the first solvation shell, is what permits the electron to localize in both bulk water and defect-containing ice.
%
%
\subsection{Excess Electron in Low Dimensional Solvation Environments}

%After validating our methodology in bulk water and exploring the structural effects of excess $e^-$ in ice, we next examine how confinement and dimensionality of solvation influence the electron localization.
%
Now we look at low dimensional solvation structures wherein the dangling bonds are always present, while the orientation and concentration of these bonds are highly dynamic.
Specifically, we consider (a) a two-dimensional (2D) monolayer of water confined between two hydrophobic sheets and (b) a one-dimensional (1D) single water chain confined in a hydrophobic tube.

\begin{figure}[htbp!]
	\centering
		\includegraphics[width=\textwidth]{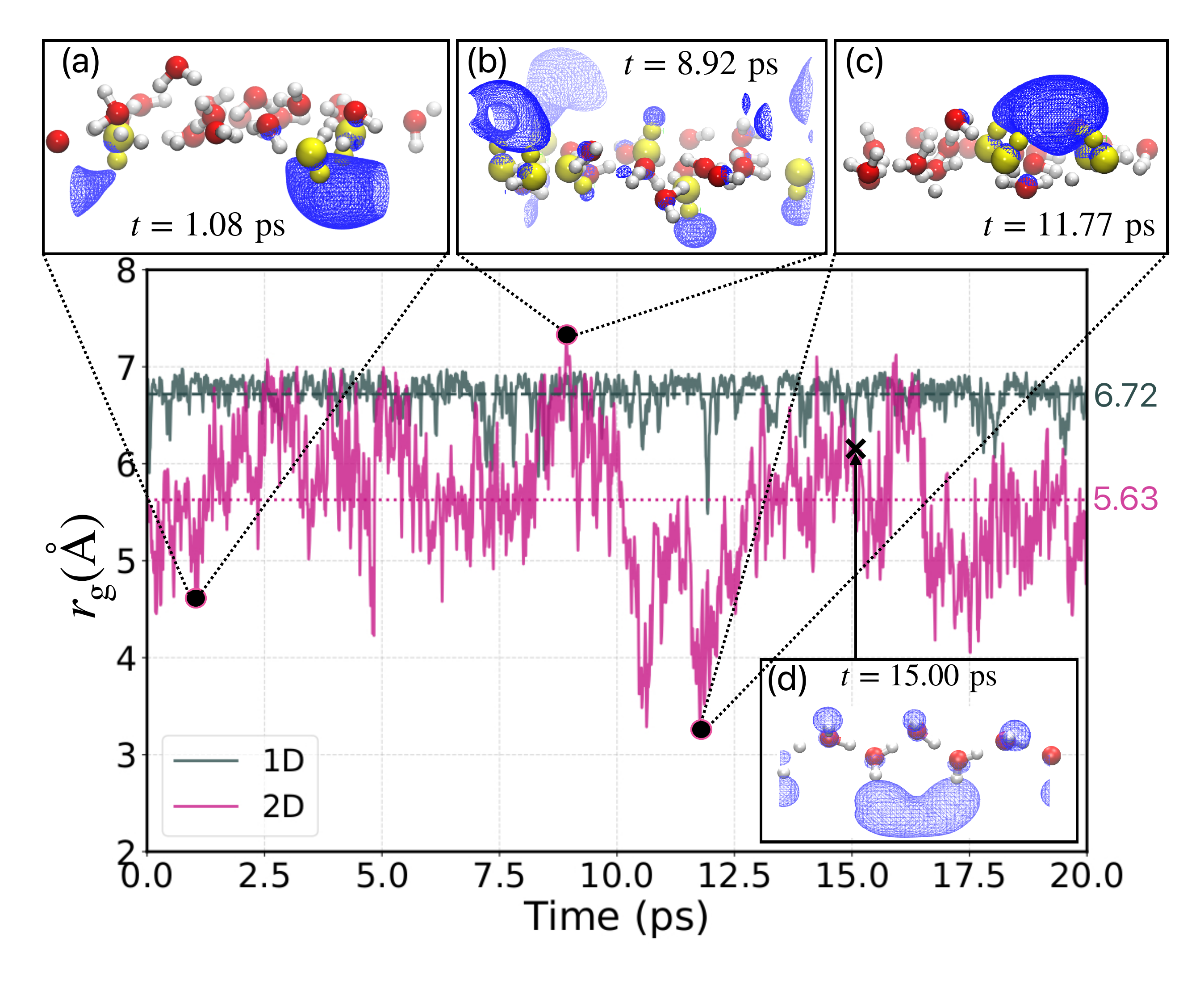}
	    \caption{Time evolution of $r_{\rm g}$ of spin density of excess $e^-$ for the 2D monolayer and 1D chain water molecules. Spin densities are shown for both cases as mesh with isovalue 0.0001~au for selected frames. 
        (a)-(c) For 2D monolayer: dangling O-H bonds interacting with $e^-$ are visualized with yellow spheres. (d) For 1D chain: spin density at $t=15.00$~ps is displayed as an example.}
	\label{f:gra_rg}
\end{figure}

%\subsubsection{Monolayer Water (2D System)}
H-bond analysis (\fref{fig:hbond_all}) revealed that in the 2D system, water molecules exhibited considerable under-coordination: 56\% of the molecules formed three H-bonds, 31\% formed four, and about 9\% formed only two.
Although theoretical studies have established that under-coordinated water molecules, especially those with dangling O-H bonds, tend to stabilize excess electrons,\cite{dipole_HE_98,pp_HE_2009,turi_Sci_05_HE,JPCA_cluster_08_HE,Angew_MP2_2019_HE,venkat_nat_comm_21_HE,Car_PCCP_24_HE,Boero_JPCA_07_HE} our simulations are revealing a highly diffused and dynamic nature.
%localization.% different picture.
%
The average $r_{\rm g}$ was relatively large at $5.63$~\AA, despite the lateral box dimensions being only $12.76 \times 12.28$~\AA$^2$ (\fref{f:gra_rg}).
%illustrates the time evolution of the  for the 2D  and 1D water systems. 
%
Although the dynamic nature of $r_{\rm g}$ of surface bound electron has also been reported by Frigato \textit{et al.}, their simulations yielded lower average values.\cite{JPCA_cluster_08_HE}
This difference is likely due to the medium-sized water clusters they studied wherein the electron can access interior or quasi-bulk regions that offer stabilizing solvation environments with at least four three-fold coordination, leading to stronger localization and thus a lower $r_{\rm g}$. 
In contrast, the monolayer arrangement in our study, combined with the hydrophobic confinement, limits such a structural flexibility and reorganization of the solvent molecules. 
Even with 21 water molecules in a periodic box, the system could not reorganize into a more bulk-like configuration capable of supporting long-lived cavity structure.
%It reveals that the 2D system has an average $r_{\rm g}$ of $5.63$~\AA, , indicating a large spatial spread of the excess electron. 
%
As a result, the electron remains largely delocalized across the plane.
%
%and showed significant fluctuations 
The significant fluctuation over the 20 ps trajectory suggests that the electron was dynamically delocalized.
%that although transient localization events occur, the spatial extent of spin density varies throughout the 20 ps trajectory. 
%

At the start of the trajectory, the spin density was localized at the bottom interface of the water layer, stabilized by three nearby water molecules, each possessing a free O–H bond directed toward the bottom surface; See \fref{f:gra_rg}(a). 
Such an orientation created a localized semi-circular cavity-like structure favorable for spatial electron accumulation. 
During this period, no other dangling O–H bonds were directed toward the interface, leading to a transient localization. 
As the simulation progressed, structural reorganization led to the appearance of multiple dangling O–H bonds pointing toward both sides of the layer; See \fref{f:gra_rg}(b). 
This created multiple sites across the plane, resulting in increased delocalization and higher $r_{\rm g}$, as illustrated in $t=8.92$~ps.

The most compact electron configuration occurred at $t = 11.77$~ps, with $r_{\rm g}$=3.3~\AA, where four dangling O–H bonds were all oriented toward the same side of the layer; See \fref{f:gra_rg}(c). 
%
%This structure was geometrically consistent with the surface-bound electron configurations reported by Turi \textit{et al.} in QCMD simulations of anionic water clusters.\cite{turi_Sci_05_HE}
%
Correlation between different frames showed that such compact configurations were short-lived and rare, reinforcing the role of dynamic molecular rearrangements in destabilizing localized states and enhancing delocalization in the monolayer of water molecules.
%confined system.
%
Rather than static stabilization by under-coordinated molecules,\cite{cluster_HE_04,cluster_HE_05} our results pointed toward a dynamic picture governed by transient structural rearrangements.\cite{PNAS_HGordon_cluster_HE}
Throughout the trajectory, hydrogen-bond connectivity was flexible, with water molecules continuously switching between different coordination states. 
Consequently, several water molecules that were under-coordinated at one moment were no longer so in subsequent frames, leading to transient formation and dissolution of dangling OH bonds. 
Such a dynamic nature directly influenced the spatial distribution of the spin density (symmetry described in Section S4, Supporting Information).

In the case of a water chain, hydrogen-bond analysis revealed that each water molecule maintained two hydrogen bonds, one donor and one acceptor, resulting in a fully extended chain with no excess bonding flexibility; See \fref{fig:hbond_all}.
Such an arrangement left one O–H bond per molecule dangling, enabling accumulation of electron density along the chain (\fref{f:gra_rg}). 
%
%On the other hand, the $r_{\rm g}$ excess $e^-$ shows a consistent spatial extent in the 1D water chain.
%
As a result, the water chain exhibited a more uniform behavior.
The average of $r_{\rm g}= 6.72$~\AA, nearly matches half of the total chain length ($12.76$~\AA), implying that the spin density spreads uniformly across the entire water chain (detailed description of shape described in Section S4, Supporting Information).
%

%Unlike in bulk water, the 2D environment does not support stable, long-lived localization.
%

\begin{figure}[ht!]
	\centering
		\includegraphics[width=\textwidth]{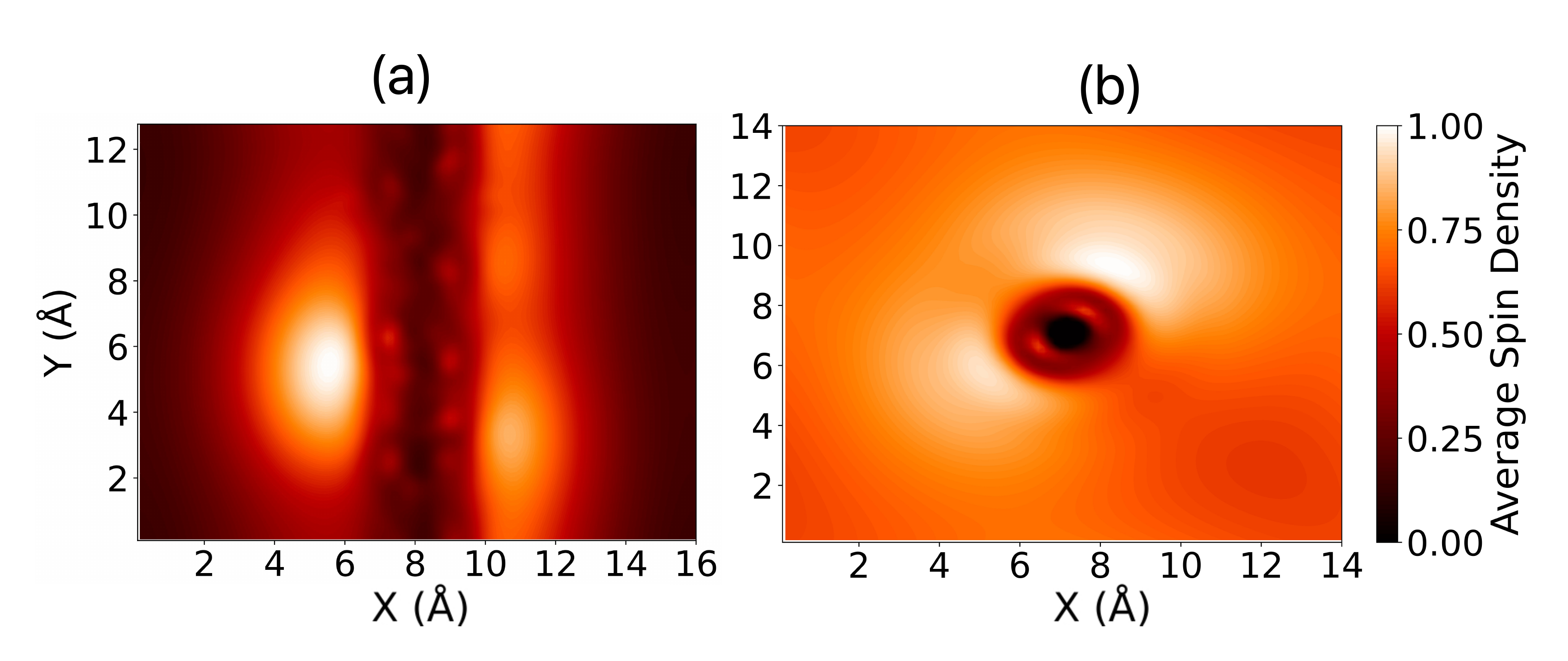}
	    \caption{Two-dimensional projection of normalized spin density on $XY$ plane for (a) 2D monolayer of water molecules confined along $YZ$ plane and (b) water chain along $Z$-axis.
        }
	\label{f:gra_cnt_proj}
\end{figure}
%The lower panel represents the projection of the average number of H-atoms along those planes for (c) a single layer of water and (d) a water chain. The orange dashed line is showing the average number of H-bonds per molecule along $X$-direction.
%

\begin{figure}[ht!]
	\centering
		\includegraphics[width=0.9\textwidth]{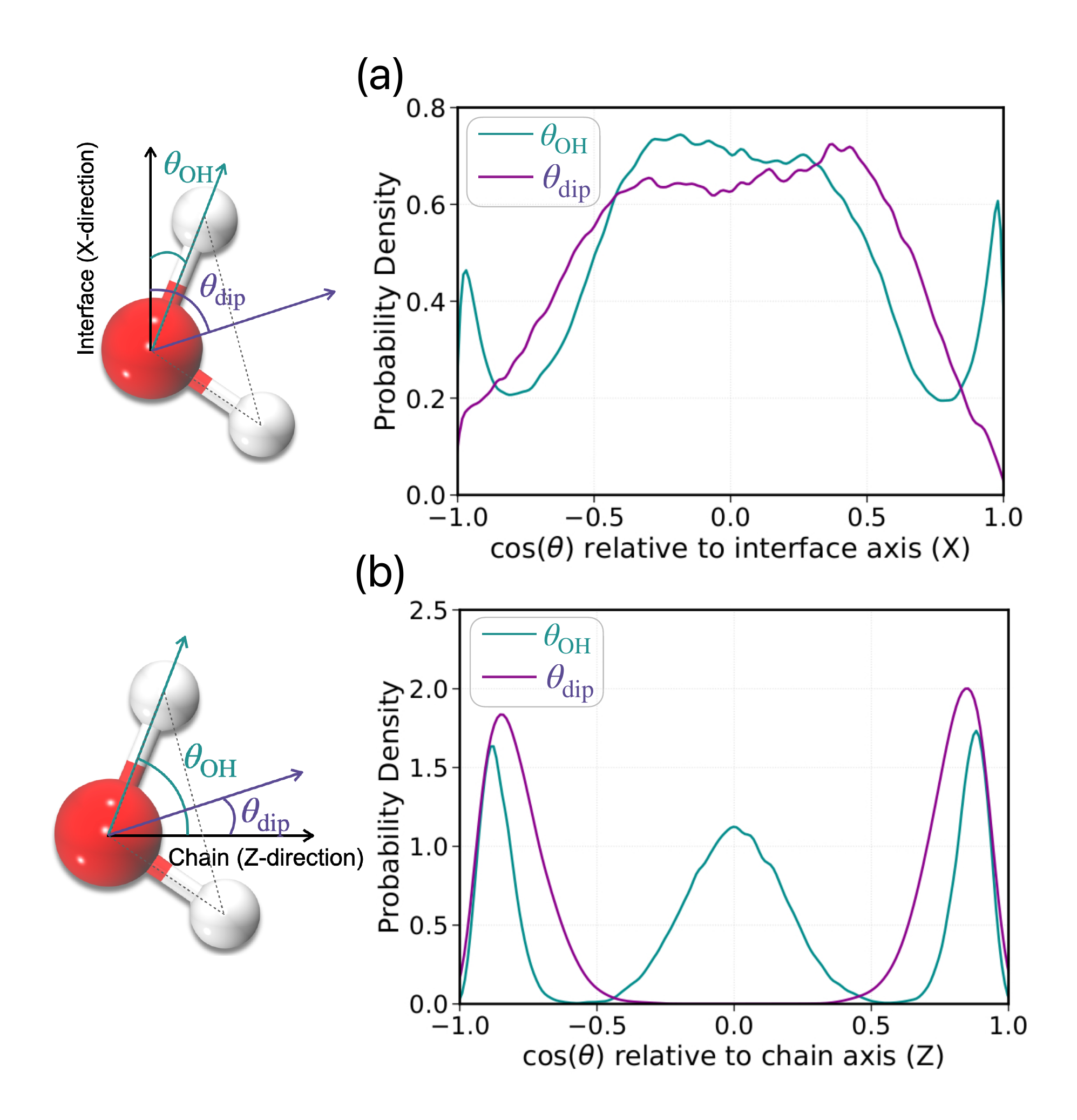}
	    \caption{Average distribution of OH vector and dipole angle for (a) 2D system with respect to the interface axis ($X$-direction) (b) 1D water chain with respect to the chain axis ($Z$-direction).
        }
	\label{fig:oh_dip_1d_2d}
\end{figure}

To gain deeper insights into the spatial nature of the excess electron in low-dimensional environments, we performed a time-averaged projection analysis of the spin density on the $XY$ plane for both the monolayer and the water chain (\fref{f:gra_cnt_proj}a-b), where the plane of the monolayer of water molecules is $YZ$, and the water chain is along the $Z$-axis (See Supporting Information, Fig. S1(c)-(d)).
In the case of monolayer, the electron density is predominantly localized at the two sides of the monolayer, with negligible presence within the midplane, as shown in \fref{f:gra_cnt_proj}a. 
This vertically confined yet laterally diffused distribution directly correlates with the angular arrangement of water molecules.
As shown in \fref{fig:oh_dip_1d_2d}a, OH vectors predominantly lie within the plane or tilt moderately toward the top and bottom leaving free sites for electron occupation.
The corresponding dipole angle distribution further supports this observation, with a broad population between cosine values of $-0.5$ and $+0.5$, indicating that dipole moments reside mostly within the plane or are inclined $60$\textdegree~ with respect to the interface normal. 
Therefore, the electropositive pockets result in a fully delocalized spin density across the $XY$ plane.
%

%Such a bonding environment leaves many dangling OH groups outward, forming a delocalized electron distribution.

%
The water chain exhibited a cylindrically symmetric shell-like spin density around the chain axis ($Z$-direction), with noticeable depletion along the central core (\fref{f:gra_cnt_proj}b).
This delocalization is consistent with the constrained molecular arrangement in the 1D channel.
The OH orientation analysis with respect to the chain axis (\fref{fig:oh_dip_1d_2d}b) reveals that each water molecule orients one OH bond radially outward ($\cos \theta= \pm 1$), while the other remains aligned ($\cos \theta=0$) with the chain axis to maintain hydrogen bonding continuity.
%while with a large fraction remaining non-hydrogen-bonded due to severe spatial constraints. 
%
Simultaneously, the water dipoles are found to be almost entirely directed along the chain, as indicated by cosine values restricted between $\pm 0.5 ~{\rm to}~ \pm 1.0$. 
These radially outward OH groups form shallow and spatially distributed electropositive regions, allowing the spin density to delocalize evenly along the length of the chain. 
However, due to the surrounding hydrophobic confinement, the six dangling O–H bonds could not reorient collectively to form a compact electrostatic cage, as observed transiently in the monolayer case, preventing any spatial or temporal localization of the electron.
%
%Unlike the 2D system, this rigid hydrogen-bond network and symmetrical OH orientation preclude the formation of localized trapping sites, explaining the large and stable $r_{\rm g}$ observed in the 1D case.

%
%As a result, the electron avoids these zones and instead adopts a toroidal distribution with high mobility around the chain. 
%
%These behaviors imply that collective dipole orientation and H-bonding play a decisive role in governing electron localization.
\section{Discussions and Conclusions}

\begin{figure}[h!]
	\centering
		\includegraphics[width=\textwidth]{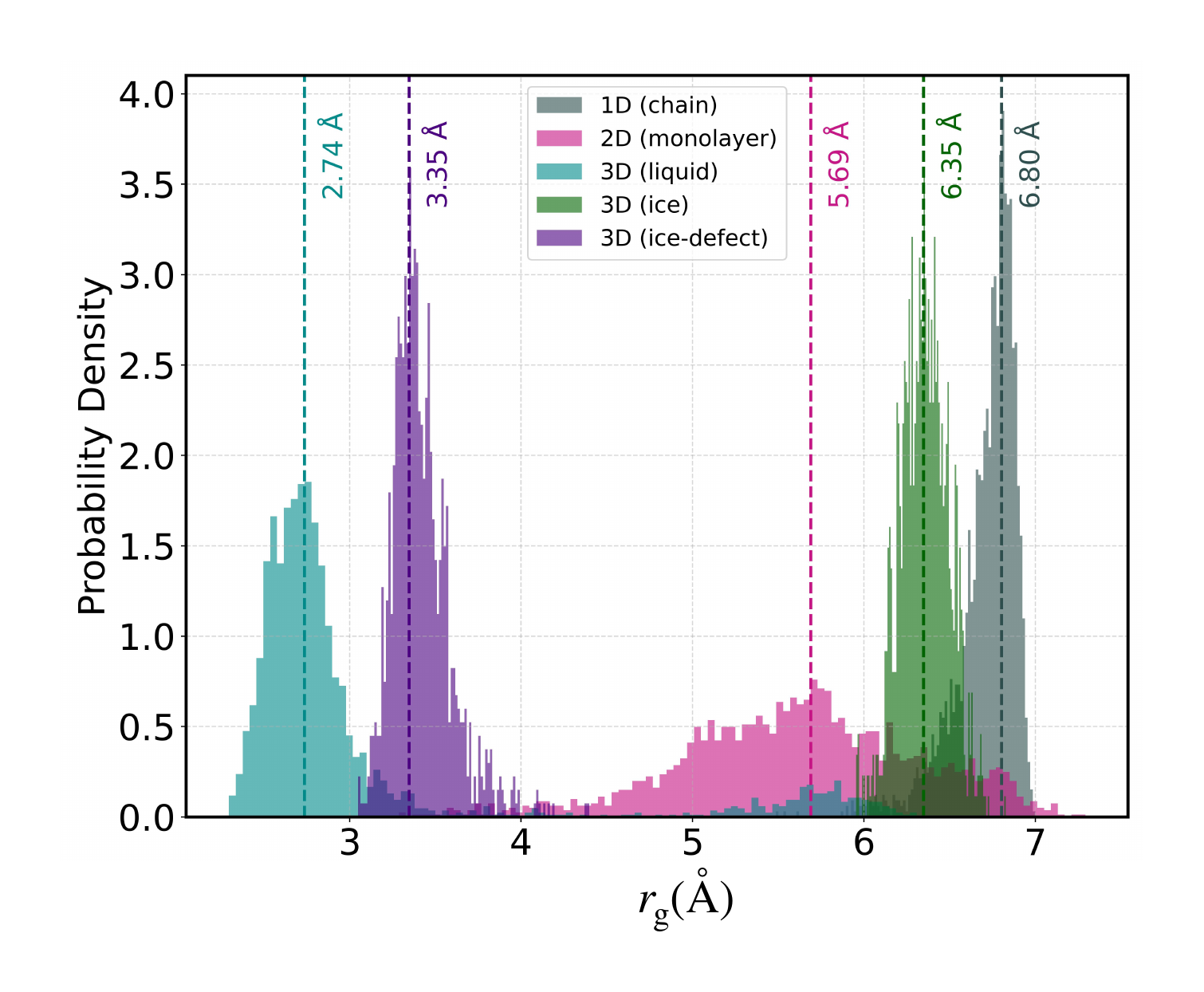}
	    \caption{Distribution of $r_{\rm g}$ for 3D (liquid, ice; with and without defect), 2D monolayer and 1D water chain systems. The most probable values of $r_{\rm g}$ are mentioned in the plot.}
	\label{fig:td}
\end{figure}

The solvation of an excess electron in water is fundamentally governed by a delicate balance between H-bond network topology, molecular orientation, and structural flexibility. 
%
%Across the transition from bulk liquid (3D) water to confined 2D monolayer and 1D water chain, we observe profound changes in electron localization behavior, driven by the geometric and dynamic constraints imposed by dimensionality reduction.
%
%By systematically analyzing the contrasting environments, from bulk liquid water (3D) to confined 2D monolayer and 1D water chain, including crystalline and defective ice, we can now outline the nature of solvated $e^-$ and understand what controls the electron stability in aqueous systems.
%
%The key to excess‑electron localization is not merely creating a first‑shell cavity, but also preserving an extended hydrogen‑bond network beyond it. 
In bulk liquid water, the introduction of the electron locally creates a cavity with low spatial extent of spherical spin density ($\langle r_{\rm g}\rangle=2.7$~\AA, \fref{fig:td}).
It disrupts four H‑bonds around the electron position, reducing first‑shell water to three coordinated on average.
The long-lived localized state is formed with the support of the second solvation shell, which remains fully with four hydrogen‑bonds per molecule, as shown in \fref{fig:hbond_bulk}. 
In defect‑free ice, by contrast, the rigid, undisturbed tetrahedral lattice, with all four coordinated water molecules (\fref{fig:hbond_all}), does not accommodate any local bond breakage, and thus no cavity can be formed. 
As a result, the electron remains delocalized ($\langle r_{\rm g}\rangle=6.34$~\AA, \fref{fig:td}).
%
%Introducing a molecular vacancy creates local disruption, by forming three-coordinated water molecules (\fref{fig:td}b).
%while the surrounding ice lattice beyond those immediate neighbors remains four coordinated. 
%
The structural relaxation around the cavity was found to be only minor, and the dangling OH bonds directed towards the cavity, where the electron was trapped ($\langle r_{\rm g}\rangle=3.4$~\AA).
%
%Due to the absence of dynamic solvent rearrangement at 77~K, the extended tetrahedral structure cannot be completely restored as liquid water.
%
Therefore, we can conclude that the presence of dangling OH bonds, all  
pointed towards a small spatial region
%is a minimal yet 
is a necessary requirement for electron localization.
%
%The slightly elevated $r_{\rm g}$ in defect-containing ice, compared to liquid water, arises from the rigidity of the ice lattice, which restricts the reorganization of nearby OH groups and hinders the formation of a compact, spherical solvation cage.
%$

Systems containing monolayer and chain of confined water molecules have under-coordinated water molecules and the dangling OH bonds are sufficiently present (\fref{fig:hbond_all}).
However, a dynamically stable localized structure of excess electron is not found in these systems; Notice $\langle r_{\rm g}\rangle=5.6\pm0.69$~\AA~ for water monolayer, and $6.7\pm 0.18$~{\AA} for water chain (\fref{fig:td}).
The significant spread of spin density in the case of monolayer is a result of the spread of the dangling O-H bonds and the respective dipole orientations along the interface.
%can facilitate the accumulation of electron density.
%
Due to the absence of a dynamically stable second solvation shell, the orientation and the location of the dangling OH bonds changed rapidly.
%
%Change in structure is also evident from the spread of $q$  in \fref{fig:td}b).
%
This has resulted in a dynamically changing structure of the spin density.
%
%
%Compared to the bulk water, where the second shell compensates and reinforces the dipolar field of the first‑shell waters, locking in the cavity, in case of 2D arrangements, adjacent molecules constantly reorient, driven by their own dipole‑minimization\cite{dipole_HE_98} and incomplete H-bond network constraints.
%
%Therefore, it leads to rapid flipping of local electrostatic basins and sustained delocalization.
%

The situation is more extreme in the case of a water chain.
%where six water molecules are present whch
%According to the coordination number of electron in liquid water, typically involving 3–5 water molecules in the first solvation shell, the 1D chain likewise contains this same number of waters, yet its rigid, 
Here, the linear arrangement prevents them from reorienting to form a stabilizing cavity around the electron.
Like in the case of the monolayer system, the water molecules in a chain do not cluster or spread laterally to build a coordination sphere that is needed for forming a cavity or a localized spin density. 
The electron remains uniformly delocalized ($\langle r_{\rm g}\rangle=6.7$~\AA) with no sign of temporal localization.

Thus, we also conclude that presence and stability of the  second solvation shell is  a necessary condition for the dynamic stabilization of the hydrated electron. 
\begin{acknowledgement}
{The authors thank Dr. Sagarmoy Mandal for helpful discussion.} % and
%Dr. Tobias Kl\"offel (FAU, Germany) for helpful discussions.}
%
Financial support from National Supercomputing Mission (NSM) (Subgroup Materials and Computational Chemistry) and Science and Engineering Research Board (India) under the MATRICS (Ref. No. MTR/2019/000359) are gratefully acknowledged.
R.K. thanks the Council of Scientific \& Industrial Research (CSIR), India for her PhD fellowship and IIT Kanpur for the FARE fellowship.
Computational resources were provided by Param Sanganak (IIT Kanpur), and Param Utkarsh (CDAC, Bangalore) under the National Supercomputing Mission, India.

\end{acknowledgement}

%%%%%%%%%%%%%%%%%%%%%%%%%%%%%%%%%%%%%%%%%%%%%%%%%%%%%%%%%%%%%%%%%%%%%
%\appendix
%\input{SI/main}
%\begin{suppinfo}

%{The Supporting Information is available free of charge on the ACS Publications website.}
%

%The Supporting Information contains: 
%Supporting information contains: (i) comparison of RDF for 32 water system with different SIN(R) parameters, (ii) snapshots of systems used for benchmark performance, (iii) detail of WS-MTD parameters for formamide hydrolysis, (iv) description of the least square error calculation, (v) different contributions to total computing time per MD step for 32 water system, and (vi) different contributions to total computing time per MD step for QM/MM calculation.
%

%\end{suppinfo}

\bibliography{achemso-demo}



\section*{Supplementary material}

# Supporting Information: Nature of Hydrated Electron in Varied Solvation Environments

Ritama Kar[1] and Nisanth N. Nair[1,*]

[1]*Department of Chemistry,*

*Indian Institute of Technology Kanpur (IITK),*

*Kanpur - 208016, India*

_____

\* nnair@iitk.ac.in

# S1. Supercell Used for Modeling Systems with Ice, Monolayer and Chain of Water Molecules

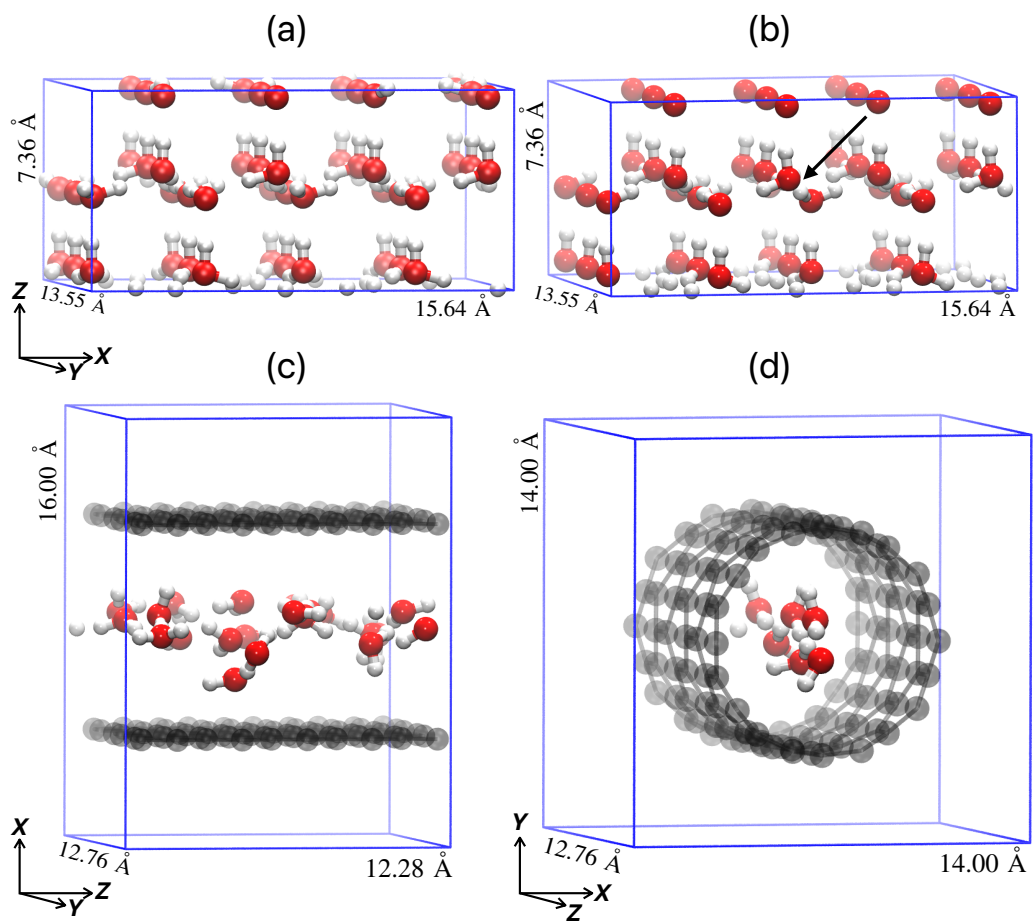

FIG. S1. (a) Perfect hexagonal ice and (b) with a molecular defect in a periodic box. The defect site is pointed with the black arrow. (c) Two-dimensional monolayer water in periodic box; (d) One-dimensional water chain inside periodic box. Oxygen atoms are shown in red and hydrogens are white. The dummy atoms are in black. The box dimensions are mentioned.

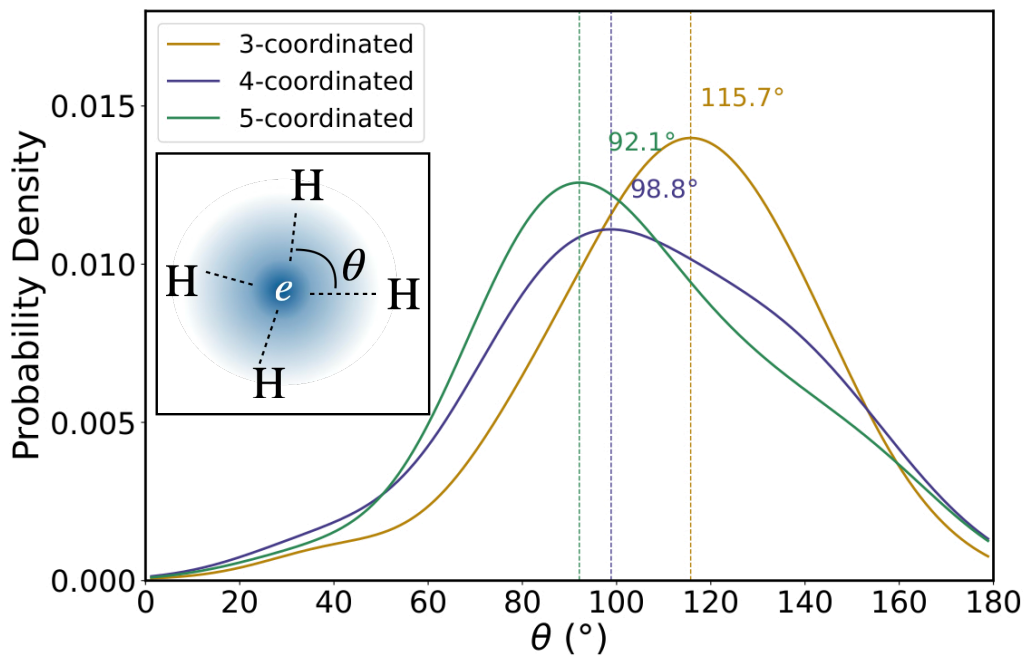

FIG. S2. Angular distribution of H-$e$-H for various coordination numbers is presented, with the maximum probable values labelled; a graphical illustration of the angle for four coordinated $e^-$ is in the inset.

## S2. Angular Distribution of Water Molecules Around Electron

We analyzed the H–$e$–H angle distribution for different coordination numbers, as shown in Figure S2. The inset of Figure S2 illustrates a four-hydrogen configuration, which shows the maximum peak at ∼99°, roughly resembling a near-tetrahedral geometry centered on the electron. A three-hydrogen coordination peaks at ∼92°, indicative of a trigonal arrangement, while five-coordination results in a peak at ∼116°, consistent with a trigonal-bipyramidal geometry. This results matches well with the previous theoretical findings.[1, 2]

## S3. Nature of H-bonds of First Solvation Shell Water Molecules

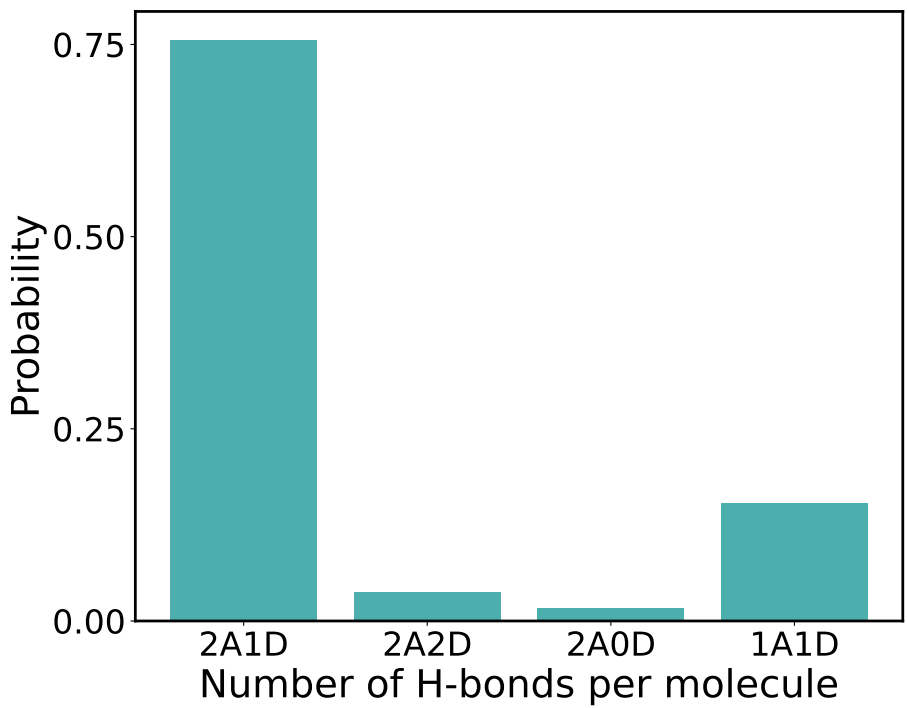

FIG. S3. Probability of H-bond population with different donor-acceptor nature in the first solvation shell of liquid water. Here $m$A$n$D implies $m$ number of acceptor bonds and $n$ number of donor H-bonds.

Our result confirmed that the most common H-bonding configuration is two acceptors and one donor H-bond (2A1D), with a marked decrease in the 2A2D configuration typically seen in bulk water. The reduction in donor H-bonds is therefore attributed to the dangling O-H bond, which is interacting with the excess electron.

## S4. Measurement of Asphericity

Asphericity is the order parameter to characterize the spherical symmetry.[3] The value close to 0 corresponds to a perfect spherical structure, while 1 represents total asymmetry. This quantity ($A$) is calculated from the gyration tensor as explained in the Methods section of the article, Equation 2, using:

$$\langle A \rangle = \frac{(\langle \mathrm{Tr}^2 - 3\mathrm{M} \rangle)}{\langle \mathrm{Tr}^2 \rangle} \ , \tag{1}$$

with the eigenvalues obtained from the **S** tensor. Tr and M are defined as:

$$\mathrm{Tr} = R_1^2 + R_2^2 + R_3^2$$
$$\mathrm{M} = R_1^2 R_2^2 + R_1^2 R_3^2 + R_2^2 R_3^2 \ . \tag{2}$$

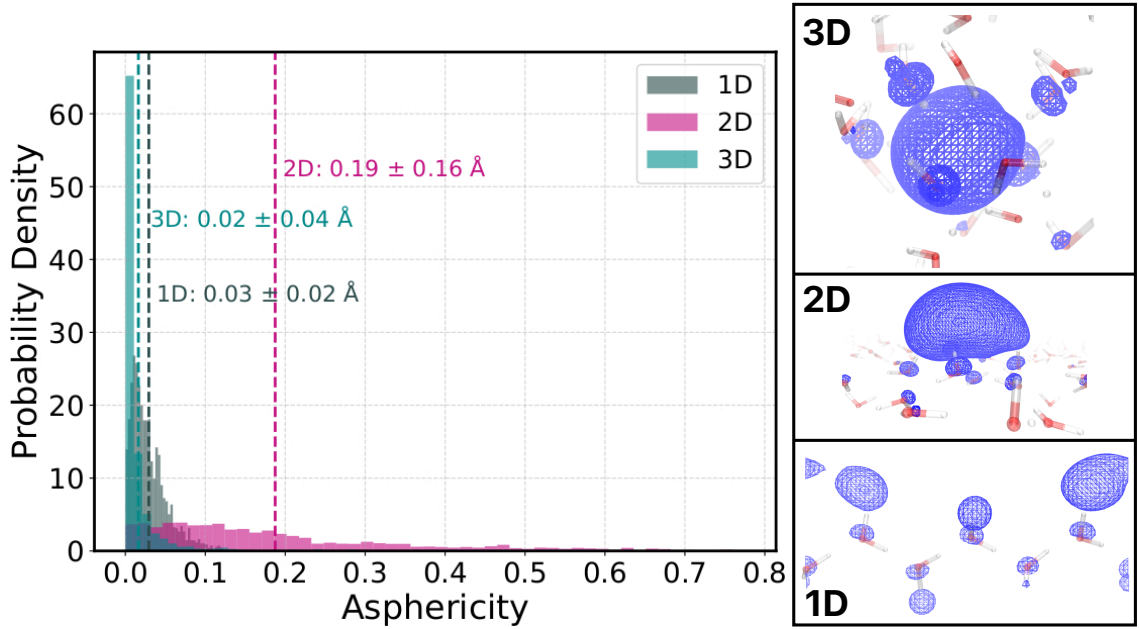

FIG. S4. Average distribution of asphericity of excess electron in bulk liquid water (3D), monolayer (2D), and water chain (1D) systems. The average value with standard deviation is mentioned in the plot. Representative snapshots with spin density are shown as mesh having isovalues of 0.001 (3D), 0.0003 (2D), and 0.0001 (1D) a.u.

We have calculated this parameter for 2000 frames with 10 fs apart for 3D, 2D, and 1D water systems. As shown in Figure 3 in the article, 3D system shows a perfect spherical

arrangement of spin density, with the asphericity value maximized at 0.0. Monolayer water arrangement allows the spin distribution with 2D alignment, with a large standard deviation, suggesting the symmetric spherical structure is not stable. On the other hand, water chain shows a distribution closer to 0, explaing the spherical charge droplet on every dangling OH group across the chain. Spin densities of different systems are visualized using different isovalues, as the electron density in 2D monolayer and 1D chain is highly delocalized, lacking the concentrated regions of density (isovalue=0.001 a.u.) observed in 3D liquid water.



\end{document}